\documentclass[useAMS,usegraphicx]{mn2e}
\usepackage{times,amssymb,hyperref,subfigure,epsfig}

\begin{document}

\title[Collisional evolution of eccentric planetesimal swarms]
  {Collisional evolution of eccentric planetesimal swarms}

\author[M. C. Wyatt et al.]
  {M. C. Wyatt$^1$\thanks{Email: wyatt@ast.cam.ac.uk},
   M. Booth$^1$,
   M. J. Payne$^1$,
   L. J. Churcher$^1$\\
  $^1$ Institute of Astronomy, University of Cambridge, Madingley Road,
  Cambridge CB3 0HA, UK}

\maketitle

\begin{abstract}
Models for the steady state collisional evolution of low eccentricity planetesimal
belts identify debris disks with hot dust at 1AU, like $\eta$ Corvi and HD69830, as
anomalous since collisional processing should have removed most of the planetesimal
mass over their $>1$ Gyr lifetimes.
This paper looks at the effect of large planetesimal eccentricities ($e \gg 0.3$)
on their collisional lifetime and the amount of mass that can remain at late times 
$M_{\rm{late}}$.
Assuming an axisymmetric planetesimal disk with common pericentre distances and
eccentricities $e$, we find that $M_{\rm{late}} \propto e^{-5/3}(1+e)^{4/3}(1-e)^{-3}$.
For a scattered disk-like population (i.e., with common pericentre distances but range of
eccentricities), in the absence of dynamical evolution, the mass evolution at late
times would be as if only planetesimals with the largest eccentricity were present
in the disk.
Despite the increased remaining mass, higher eccentricities do not increase
the amount of hot emission from the collisional
cascade until $e>0.99$, partly because most collisions occur near
pericentre thus increasing the dust blow-out diameter.
However, at high eccentricities ($e>0.97$) the blow-out population extending
outwards from pericentre may be detectable above the collisional cascade;
higher eccentricities also increase the probability of witnessing a recent collision.
All of the imaging and spectroscopic constraints for $\eta$ Corvi can be explained
with a single planetesimal population with pericentre at 0.75AU, apocentre at 150AU,
and mass $5M_\oplus$;
however, the origin of such a high eccentricity population remains challenging.
The mid-infrared excess to HD69830 can be explained by the ongoing
destruction of a debris belt produced in a recent collision in an eccentric
planetesimal belt, but the lack of far-infrared emission would require small
bound grains to be absent from the parent planetesimal belt, possibly due to 
sublimation.
The model presented here is applicable wherever non-negligible planetesimal
eccentricities are implicated and can be readily incorporated into N-body
simulations.
\end{abstract}


%%%%%%%%%%%%%%%%%%%%%%%%%%%%%%%%%%%%%%%%%%%%%%
%%%%%%%%%%%%%%%%%%%%%%%%%%%%%%%%%%%%%%%%%%%%%%
\section{Introduction}
\label{s:intro}
A natural byproduct of the planet formation process, at least
in the core accretion model, is the formation of planetesimals
(Lissauer 1993).
Evidence for planetesimals following the protoplanetary disk phase
comes from debris disks, a phenomenon in which main sequence stars
exhibit an infrared excess which is attributed to the thermal emission
of dust released from planetesimals in collisions and sublimation
(see review in Wyatt 2008).
The Solar System has its own debris disk, the majority of which is
in the asteroid and Kuiper belts.

Typically extrasolar debris disks are observed to lie in a ring at
a single radius (Greaves et al. 2005; Kalas, Graham \& Clampin 2005; Schneider et al.
2009), or where they are not imaged the emission spectrum is dominated by a single
temperature (Chen et al. 2006).
This motivates considering these disks as planetesimal belts that are
directly analogous to the asteroid and Kuiper belts (Moro-Mart\'{i}n et al. 2008),
and the disks where dust is detected at multiple radii
(Wyatt et al. 2005; Absil et al. 2006; Smith et al. 2009a;
Backman et al. 2009; Chen et al. 2009) are usually inferred to have multiple planetesimal belts.
In the absence of other dynamical processes, the evolution of these belts
is expected to be dominated by collisions which grind away the mass of
the largest objects into dust which is subsequently removed by radiation
pressure (or P-R drag in the case of the Solar System) (Wyatt 2009).

The collisional evolution of the planetesimal belts of the Solar System has
been studied extensively.
Collision rates can be derived accurately between objects moving on 
given orbits (\"{O}pik 1951; Wetherill 1967; Greenberg 1982;
Bottke et al. 1994; Vedder 1996; Dell'Oro \& Paolicchi 1998),
and the steady state size distribution of the belts resulting from
their collisional evolution is both well understood analytically
(Dohnanyi 1969; Tanaka et al. 1996; O'Brien \& Greenberg 2003; Kobayashi
\& Tanaka 2009) and one-dimensional numerical models
of this evolution that include a realistic prescription for the outcome
of collisions provide a good fit to the observed size distributions in the
asteroid belt (Davis et al. 1989; Durda, Greenberg \& Jedicke 1998;
Bottke et al. 2005; O'Brien \& Greenberg 2005) and
Kuiper belt (Stern \& Colwell 1997; Davis \& Farinella 1997;
Kenyon \& Bromley 2004).

The approach to considering the collisional evolution of extrasolar debris disks
is slightly different in that the orbital element and size distributions of the
parent bodies are poorly constrained, rather it is important to generalize the effect 
of this evolution on debris disk observability with respect to parameters such as 
initial planetesimal belt mass, radius and mean eccentricity.
Analytical models that achieve this were developed by Dominik \& Decin (2003)
who considered the evolution of a mono-disperse
planetesimal belt (i.e., with planetesimals all of the same size) that feeds a
population of smaller planetesimals and dust that is observed.
This model was later refined by Wyatt et al. (2007a) to consider the parent
planetesimals and the smaller objects to form a continuous size distribution
defined by a single power law as expected for the steady state case where planetesimal
strength is independent of size (Dohnanyi 1969; Tanaka et al. 1996).
A size dependent planetesimal strength was later included in such models
by L\"{o}hne et al. (2008) resulting in a more realistic 3 phase size distribution
(e.g., O'Brien \& Greenberg 2003).
Both the Wyatt et al. (2007b) and L\"{o}hne et al. (2008) models were applied to the
statistics of detections of debris disks around A stars and Sun-like stars to
show that these could be explained if the majority of such debris disks evolve
purely due to steady state collisional evolution.

One important result that came out of the Wyatt et al. (2007a) study was the concept
of a maximum planetesimal belt mass, and hence a maximum dust luminosity, that can
remain for a given radius belt at a given time, regardless of its initial mass.
Although this is no longer strictly true when a size dependent strength
is used, L\"{o}hne et al. (2008) showed that initial mass has a relatively modest
effect on the mass remaining at late times, and concurred that for realistic planetesimal
belt parameters there is indeed a maximum planetesimal belt mass and dust
luminosity for a given age (see also Heng \& Tremaine 2009).
This concept was used by Wyatt et al. (2007a) to show that 1-2Gyr systems
like $\eta$ Corvi and HD69830 that have large quantities of hot dust at
$\sim 1$AU (Wyatt et al. 2005; Beichman et al. 2005; Smith, Wyatt \& Dent 2008),
cannot be replenishing that dust from planetesimal
belts that are coincident with the dust (i.e., analogous asteroid belts).
They concluded that the parent bodies of the observed
dust must have originated at larger radii ($\gg$ several AU) where collisional
processing times would have been longer.
The paper also concluded that the hot dust is transient and proposed that this might
have been scattered in from an outer belt in an epoch akin to the Late Heavy
Bombardment in the Solar System (see review in Hartmann et al. 2000).
There are now several examples of systems exhibiting hot dust that appears to be
transient by the criterion described by Wyatt et al. (2007a)
(e.g., di Folco et al. 2007; Akeson et al. 2009; Mo\'{o}r et al. 2009).

The motivation of this paper is to consider whether it is possible to circumvent the 
conclusion that the hot dust in systems like $\eta$ Corvi and HD69830
must be transient by postulating a population of parent planetesimals on highly
eccentric orbits ($e \gg 0.3$).
In such a model the hot dust would originate from material close to pericentre,
and the parent population could be long-lived because the planetesimals spend most of
their time at apocentre.
This would challenge our traditional view of debris disks as belts of planetesimals
with modest eccentricity ($e<0.3$), which is also implicit in the models of Wyatt et 
al. (2007a) where collision velocities are assumed to be proportional to Keplerian 
velocity times a mean eccentricity for the belt, and in the
models of L\"{o}hne et al. (2008) where eccentricities up to 0.35 were considered.
However, it is clear from the Solar System that there are also populations
of planetesimals on highly eccentric orbits ($e \gg 0.3$):
the comets scattered in from the Kuiper belt (Duncan 2008);
the scattered disk of the Kuiper belt (which may be primordial in origin
and extends all the way to the Oort cloud) (Gomes et al. 2008);
as well as the Near Earth Asteroids (Bottke et al. 2002).
While the contribution of these populations to the dust content of the
zodiacal cloud may be small, the cometary contribution could be as much
as 90\% (Ipatov et al. 2008; Nesvorn\'{y} et al. 2009), and may have been
significantly higher in the past, e.g., during the epoch known as the Late Heavy Bombardment
(Gomes et al. 2005; Booth et al. 2009).
Furthermore the opposite may be true for planetary systems with different architectures 
and formation scenarios, in which eccentric planetesimals may dominate.
Indeed, planet formation models often predict a highly eccentric remnant
planetesimal population (Edgar \& Artymowicz 2004; Mandell, Raymond \& Sigurdsson
2007; Payne et al. 2009).

Thus, here we develop the model of Wyatt et al. (2007a) to include interactions between
planetesimals of arbitrary eccentricities and semimajor axes (and inclinations).
Although this model does not (yet) include the more realistic assumption of a size
dependent planetesimal strength, it benefits by providing simple 
analytical formulae for collision lifetimes from which the observability of a 
planetesimal belt as a function of its eccentricity can be readily assessed.
The inclusion of a size dependent strength would be expected to give results within an 
order of magnitude of those presented here (see, e.g., Fig. 11 of L\"{o}hne et al.
2008), a level of uncertainty that is commensurate with the uncertainty in estimates 
for planetesimal strength at each size for the Solar System and for different
assumptions about planetesimal composition (see e.g., Fig. 1 of Durda et al. 1998, and 
Fig. 11 of Leinhardt \& Stewart 2009).

In \S \ref{s:2} we consider the collisional evolution of an axisymmetric disk of 
planetesimals all of which have the same pericentre and apocentre distances, and show that
the concept of a maximum remaining mass for a given age also applies in this case,
but that the remaining mass is higher for higher eccentricities (if pericentre distance is
kept constant).
The approach to calculating collision rates is similar to that of Bottke et al. (1994) in 
that we assume random mean longitudes, arguments of pericentre and longitudes of 
ascending node, but differs in using a particle-in-a-box approach to calculate the 
collision rate at a particular point on the orbit then integrating around the orbit
(as opposed to calculating this from the fraction of the orbits that the planetesimals
spend close enough that they overlap in physical space).
Our collision rate at each point also includes an integration
over the size distribution of impactors that can cause a catastrophic collision,
whereas this integration is performed after calculating the intrinsic collision
probability by Bottke et al. (1994), requiring that method to keep track of the
velocity probability distribution.
Our integration is performed using a Monte-Carlo technique, but in the case
of mutual collisions amongst a population with common eccentricities and semimajor
axes, and assuming that collision velocities are dominated by radial motion (due to
eccentricities) rather than vertical motion (due to inclinations), the collision
rate can also be derived analytically.

In \S \ref{s:4} we consider the more general situation in which planetesimals can
interact with planetesimals with different pericentre and apocentre distances, and show that
our collision rates agree with those of the most accurate studies available in
the literature.
To consider the evolution of a realistic planetesimal belt where a range of 
eccentricities and semimajor axes is present we adopt an approach similar
to Krivov et al. (2005, 2006) in that we consider the evolution of the phase space
distribution.\footnote{Note that the Krivov et al. 2005 model is not 
accurate for high eccentricities because of the way mean impact velocity 
was calculated (see \S \ref{ss:comp}).}
Here we outline a scheme for evolving the phase space distribution numerically,
and apply this to a scattered-disk like population with common pericentre
and a range of apocentre distances.
The emission properties of eccentric rings are considered in \S \ref{s:3} to assess if
the emission spectrum of real systems can be consistently explained by steady state
processing given the stellar ages.
The conclusions are given in \S \ref{s:conc}.

%%%%%%%%%%%%%%%%%%%%%%%%%%%%%%%%%%%%%%%%%%%%%%%%%%%%%%
%%%%%%%%%%%%%%%%%%%%%%%%%%%%%%%%%%%%%%%%%%%%%%%%%%%%%%
\section{Collisional lifetime of single pericentre-apocentre population}
\label{s:2}

Our approach to calculating collision rates is based on the particle-in-a-box
approach, wherein the planetesimals are assumed to be spread uniformly around an
annulus and to have a mean collision velocity (e.g., equation 28 of Wetherill 1967).
The resulting collision rate is accurate to a factor of 2 when the asteroid belt
is considered as a single annulus (Wetherill 1967).
Such techniques have also been well developed for studies of the 
accumulation of planetesimals into planets (e.g., Greenberg et al. 1978; Wetherill \& 
Stewart 1989) and it is possible to derive precise collision rates for
certain assumptions about the distributions of planetesimal eccentricities and 
inclinations, such as that these follow a Rayleigh distribution
(Greenzweig \& Lissauer 1990; Lissauer \& Stewart 1993).
However the derived collision rates are only valid where eccentricities and
inclinations are $\ll 1$.
Here we consider collision rates between pairs of orbits that can (but do not necessarily)
have significant eccentricity by splitting up the orbits into annuli, since the density and 
velocity distributions in each annulus are well defined and the particle-in-a-box approach can be 
used to work out accurate collision rates which can then be integrated around the orbit;
in this respect our approach to calculating collision rates is
similar to that of Spaute et al. (1991) which assumed $e,I \ll 1$.

%%%%%%%%%%%%%%%%%%%%%%%%%%%%%%%%%%%%%%%%%%%%%%%%%%%%%%
\subsection{Local collision rates}
Consider a planetesimal of diameter $D$ that is moving through a disk of planetesimals
with a range of sizes, where the size distribution is defined such that 
$\bar{\sigma}(D_{\rm{im}})dD_{\rm{im}}$ is the fraction of the total cross-sectional area
in the distribution that is in the size range $D_{\rm{im}}$ to $D_{\rm{im}}+dD_{\rm{im}}$.
If the local volume density of cross-sectional area of planetesimals (of all sizes) is
$\sigma_{\rm{v}}$ in AU$^2$/AU$^3$, and the relative velocity
of collisions is $v_{\rm{rel}}$ in m/s, then a particle-in-a-box calculation gives
the local rate of impacts from planetesimals in the size range $D_{\rm{im}}$ to 
$D_{\rm{im}}+dD_{\rm{im}}$ as $R_{\rm{col}}(D,D_{\rm{im}})dD_{\rm{im}}$ where
\begin{equation}
  R_{\rm{col}}(D,D_{\rm{im}}) = 2.11 \times 10^{-4} f(D,D_{\rm{im}})
                                \sigma_{\rm{v}} v_{\rm{rel}}
  \label{eq:rcol}
\end{equation}
in yr$^{-1}$, where the constant here (and in later equations) arises from the choice of
units for the various parameters, and
\begin{equation}
  f(D,D_{\rm{im}}) = \bar{\sigma}(D_{\rm{im}}) (1+D/D_{\rm{im}})^2.
  \label{eq:f}
\end{equation}
Note that due to the high relative velocities in an eccentric disk
we have ignored gravitational focussing in this collision rate
(see \S \ref{sss:wherecoll}).

The majority of the collisions suffered by the planetesimal cause negligible mass loss.
Although the cumulative effect of such cratering collisions can be important (e.g., Kobayashi
\& Tanaka 2009), here we consider the rate at which the planetesimal suffers collisions
that have sufficient energy to cause catastrophic disruption.
A catastrophic collision is defined as one in which the largest fragment remaining following 
the collision (due to both shattering and subsequent gravitational reaccumulation) has 
half the mass of the original planetesimal, and a planetesimal's dispersal threshold 
$Q_{\rm{D}}^\star$ (in J kg$^{-1}$) is defined such that catastrophic collisions are those in 
which the specific incident kinetic energy exceeds $Q_{\rm{D}}^\star$.
Dispersal thresholds have been ascertained for planetesimals of varying size and composition using
a combination of laboratory experiments and numerical modelling (e.g., Fujiwara et al. 1989; Benz 
\& Asphaug 1999).
This definition means that for a given collision velocity there
is a minimum size of object that the planetesimal must be impacted by to be catastrophically
destroyed.
Denoting this as $D_{\rm{tc}}$ gives the minimum relative size of impactor to target for
catastrophic disruption as $X_{\rm{c}}=D_{\rm{tc}}/D$ where
\begin{equation}
  X_{\rm{c}} = (2Q_{\rm{D}}^\star/v_{\rm{rel}}^2)^{1/3}.
  \label{eq:xc}
\end{equation}
Working out the rate of catastrophic collisions, $R_{\rm{cc}}$, then requires integrating
equation (\ref{eq:rcol}) from $D_{\rm{tc}}$ up to the largest object in the size distribution
\begin{equation}
  R_{\rm{cc}}(D) = 2.11 \times 10^{-4} f_{\rm{cc}}(D) \sigma_{\rm{v}} v_{\rm{rel}},
  \label{eq:rcc}
\end{equation}
where $f_{\rm{cc}}(D) = \int_{D_{\rm{tc}}(D)}^{D_{\rm{max}}} f(D,D_{\rm{im}}) dD_{\rm{im}}$.

%%%%%%%%%%%%%%%%%%%%%%%%%%%%%%%%%%%%%%%%%%%%%%%%%%%%%%
\subsection{Evolution of the size distribution}
\label{ss:evolsd}
To simplify our model for the evolution of the size distribution we make the
assumption that the distribution follows a single power law
\begin{equation}
  n(D) \propto D^{2-3q_{\rm{d}}}
  \label{eq:nd}
\end{equation}
between sizes $D_{\rm{min}}$ (in $\mu$m) and $D_{\rm{max}}$ (in km),
where the planetesimals are assumed to be spherical to get $\bar{\sigma}(D) \propto 
D^{4-3q_{\rm{d}}}$.
For the situation where planetesimal strength (i.e., the dispersal threshold)
is independent of size and the size distribution has no maximum or minimum size
it is well known that the planetesimal belt's steady state solution has a power
law size distribution with $q_{\rm{d}}=11/6$ (Dohnanyi 1969; Tanaka et al.
1996), which we use throughout this paper.
This slope arises because it means that the mass loss rate in each size bin is both
independent of size and balanced by mass gain from the fragmentation of larger
planetesimals.
A more realistic size distribution is truncated both at small sizes (e.g., due to
radiation pressure) and at large sizes (e.g., set by the largest object which formed in
the belt).
The truncation of the size distribution at the small size end causes a
ripple in the steady state size distribution (Campo-Bagatin et al. 1994).
However, the truncation at large sizes has an important longer term effect,
as it means that collisions eventually deplete the number of large objects,
since these are no longer being replenished by the destruction of yet bigger objects.

In the simple model proposed by Wyatt et al. (2007)a,
the size distribution is considered to be in quasi-steady state thus maintaining the
power law slope of equation (\ref{eq:nd}), but for mass to be lost as the largest planetesimals
are depleted on their catastrophic collision timescale.
For the size distribution of equation (\ref{eq:nd}), the factor in the collision rate for the
largest planetesimals is
\begin{equation}
  f_{\rm{cc}}(D_{\rm{max}}) = (10^{-9}D_{\rm{min}}/D_{\rm{max}})^{3q_d-5} 
     G(q_d,X_{\rm{c}}),
\end{equation}
where the assumption that $q_{\rm{d}}=11/6$ results in
$G(q,X_{\rm{c}})=0.2X_{\rm{c}}^{-2.5}+0.67X_{\rm{c}}^{-1.5}+X_{\rm{c}}^{-0.5}-1.87$
(see Wyatt et al. 2007a), which we further simplify to $G(11/6,X_{\rm{c}}) \approx 
0.2X_{\rm{c}}^{-2.5}$ which is accurate to 71\% for $X_{\rm{c}}<0.87$ and to 30\% for
$X_{\rm{c}}<0.1$.
This means that the local catastrophic collision rate of 
the largest planetesimals is
\begin{equation}
  R_{\rm{cc}}(D_{\rm{max}}) \approx 7.49 \times 10^{-10} D_{\rm{min}}^{0.5} D_{\rm{max}}^{-0.5}
                              {Q_{\rm{D}}^\star}^{-5/6}
                              \sigma_{\rm{v}} v_{\rm{rel}}^{8/3}.
  \label{eq:rccdmax}
\end{equation}

Since most of the objects which are causing this catastrophic destruction are
both large and contain most of the mass (but little of the cross-sectional
area), it is more appropriate to rewrite equation (\ref{eq:rccdmax}) using the 
following relation
\begin{equation}
  \sigma_{\rm{tot}} = 12650 M_{\rm{tot}} \rho^{-1} D_{\rm{min}}^{-0.5}
                      D_{\rm{max}}^{-0.5}
  \label{eq:stot}
\end{equation}
in AU$^2$, where $M_{\rm{tot}}$ is total mass in the distribution
in $M_\oplus$, and $\rho$ is planetesimal density in kg m$^{-3}$,
to find that
\begin{equation}
  R_{\rm{cc}}(D_{\rm{max}}) = K M_{\rm{tot}}
                              \bar{\sigma}_{\rm{v}} 
                              v_{\rm{rel}}^{8/3},
  \label{eq:rccdmax2}
\end{equation}
where
\begin{equation}
  K = 9.5 \times 10^{-6} \rho^{-1} D_{\rm{max}}^{-1} {Q_{\rm{D}}^\star}^{-5/6}
  \label{eq:k}
\end{equation}
and $\bar{\sigma}_{\rm{v}}=\sigma_{\rm{v}}/\sigma_{\rm{tot}}$ is 
the normalised volume density of cross-sectional area in AU$^{-3}$.
Later in the paper we adopt $K=2.1 \times 10^{-14}$ as a fiducial value corresponding
to $\rho=2700$ kg m$^{-3}$, $D_{\rm{max}}=2000$ km and $Q_{\rm{D}}^\star=200$ J kg$^{-1}$,
or some other equivalent combination of these parameters.

We acknowledge that the above prescription gives a highly simplistic view of the evolution
of the size distribution.
For example, for an infinite collisional cascade in which
$Q_{\rm{D}}^\star \propto D^\alpha$ the size distribution is still expected to
follow a power law, but with a slope
that depends on $\alpha$ (O'Brien \& Greenberg 2003; 
Kobayashi \& Tanaka 2009).
Since experiments have shown that $Q_{\rm{D}}^\star$
has a different slope at small sizes (in the strength regime) to that at large
sizes (in the gravity regime), and the primordial planetesimal distribution
may differ from that expected when the distribution reaches steady state, a
more realistic prescription for the size distribution has 3 power laws in
different size regimes (L\"{o}hne et al. 2008), with numerical simulations
showing that the transition from strength to gravity scaling also causes a
further wiggle in the size distribution (Durda et al. 1998).
However, despite the seemingly simplistic view of the evolution presented
here, it was found that its predictions for the evolution of mass and area in
a planetesimal belt are accurate to within an order of magnitude of more
detailed models (see, e.g., Fig. 11 of L\"{o}hne et al. 2008). 
Given that there are large uncertainties in the various parameters that
make up the constant $K$ (equation~\ref{eq:k}) --- e.g., estimates 
for planetesimal strength at a given size vary by several orders of magnitude
both from constraints from the Solar System's debris belts and for different
assumptions about planetesimal composition (see e.g., Fig. 1 of Durda et al. 1998, and 
Fig. 11 of Leinhardt \& Stewart 2009) --- we consider that this model
is sufficient to assess the impact of planetesimal eccentricity on the
observability of a debris disk, but note that it is possible to construct
a more realistic (and more complex) prescription for the size distribution 
if assumptions about how planetesimal strength varies as a function of 
size are made.

%%%%%%%%%%%%%%%%%%%%%%%%%%%%%%%%%%%%%%%%%%%%%%%%%%%%%%
\subsection{Low eccentricity approximation}
\label{ss:lowe}
Collision rates calculated using particle-in-a-box methods such as that described above 
typically use what we will call the low eccentricity approximation, since the assumptions
break down when $e \gg 0.3$.
When applied to a planetesimal belt such as the asteroid belt (e.g., Wetherill 1967),
it is assumed that the planetesimals are spread uniformly throughout a torus of radius $r$, width 
$dr$ and vertical extent $2I_{\rm{max}}r$, so that normalised cross-sectional area density can 
be approximated at all locations within the torus by
\begin{equation}
  \bar{\sigma}_{\rm{v}} = (4\pi r^2 dr I_{\rm{max}})^{-1}.
\end{equation} 
Second it is assumed that, although collisions in a realistic planetesimal belt occur at a
range of relative velocities, collision rates can be calculated using a mean
relative velocity;
this can either be calculated by considering the observed
distribution of orbital elements (i.e., for the asteroid belt), or where this is not
known by assuming that it is proportional to the Keplerian velocity at $r$,
\begin{equation}
  v_{\rm{k}}(r) = 2.98 \times 10^4 M_\star^{0.5} r^{-0.5},
\end{equation}
in m s$^{-1}$, where $M_\star$ is in $M_\odot$ and $r$ in AU,
through the mean eccentricities and inclinations of planetesimals' orbits so that
\begin{equation}
  v_{\rm{rel}}/v_{\rm{k}} = f(e,I).
\end{equation}
It can be shown that for Rayleigh distributions of eccentricities and inclinations, the mean 
relative velocities between planetesimals is given by $f(e,I)=e \sqrt{1.25+(I/e)^2}$
(equation 17 of Lissauer \& Stewart 1993).
These two assumptions can be fed directly into equation (\ref{eq:rccdmax2}) to get an
expression for the collision rate which in this approximation is the same for
the largest planetesimals at all locations within the torus:
\begin{eqnarray}
  R_{\rm{cc}}(D_{\rm{max}}) & = & 1.9 \times 10^6 \rho^{-1} D_{\rm{max}}^{-1} 
             {Q_{\rm{D}}^\star}^{-5/6} M_{\rm{tot}}
     \times \nonumber \\ & &
        r^{-13/3} (dr/r)^{-1} M_\star^{4/3} I_{\rm{max}}^{-1} e^{8/3},
\end{eqnarray}
where it was further assumed that $I=e$.
%Further assuming that $I_{\rm{max}}=I=e$ and $\rho=2700$ kg m$^{-3}$
%results in an expression that can be directly compared with equation 16 in
%Wyatt (2008).

%%%%%%%%%%%%%%%%%%%%%%%%%%%%%%%%%%%%%%%%%%%%%%%%%%%%%%
\subsection{Higher eccentricities}
\label{ss:highe}
For higher eccentricities ($e \gg 0.3$) the assumptions of \S \ref{ss:lowe} break down,
because it is expected that both the cross-sectional area density
and the relative velocity of collisions are significantly different at different
locations within the torus and so vary around the planetesimals' orbits.
To calculate the collision rate between planetesimals on eccentric orbits we use
an approach that differs from that pioneered by \"{O}pik (1951), but show in \S \ref{ss:comp}
that the two approaches get identical results.
The method is based on the particle-in-a-box approach in that it assumes that 
equations (\ref{eq:rcc}) and (\ref{eq:rccdmax2}) provide good estimates for
the planetesimal's local collision rates as long as the volume density and
relative velocity of collisions at that location are well known.
For the orbital element distributions we consider, we expect these
to be well characterised and to depend only on radius (and latitude),
and so the collision rate can be calculated by integrating the local rate 
along the planetesimal's orbit.

To both work out the local collision rate and to perform this integration a Monte
Carlo approach is used, wherein $N$ planetesimals are chosen with orbital elements chosen randomly 
from given ranges.
This section considers a disk of planetesimals all with
pericentre distances in the range $q-dq/2$ to $q+dq/2$ and apocentre distances in the range 
$Q-dQ/2$
to $Q+dQ/2$,
where $dq$ and $dQ$ are small enough to have no effect on the result, but are
kept finite for practical reasons and to allow easy implementation in
the model of \S \ref{s:4}.
The disk is assumed to be axisymmetric so that pericentre orientation $\varpi$ is
chosen randomly from the range 0 to $2\pi$, as is mean anomaly $M$.
The orbital elements for these $N$ planetesimals are then converted into 2-dimensional
positions and velocities.

%%%%%%%%%%%%%%%%%%%%%%%%%%%%%%%%%%%%%%%%%%%%%%%%%%%%%%
\subsubsection{2D approximation}
\label{sss:2dapprox}
The 3-dimensional structure of the disk is accounted for in this section by
assuming that vertical motion has little
effect on relative velocities which can thus be calculated in 2-dimensions, but 
that the vertical structure does affect the local cross-sectional area density
which is accounted for by assuming a vertical extent of $2rI_{\rm{max}}$.
Given the axisymmetric structure of the disk, the collision rate should be the same
for planetesimals at the same radius.
Thus the disk was split into a number of annuli, $N_{\rm{ann, r}}$, logarithmically
spaced in radius between $q-dq/2$ and $Q+dQ/2$.

The planetesimal positions were then used to derive $\bar{\sigma}(r)$
which is both the fraction of the total cross-sectional area in the annulus at $r \pm dr/2$,
and the fraction of time spent by planetesimals in the different annuli as they
go around their orbit.
Further dividing by the volume of the annulus, $dV_{\rm{2d}} = 4\pi r^3 (dr/r) I_{\rm{max}}$
gives the normalised volume density $\bar{\sigma}_{\rm{v}}(r)$.
For each planetesimal, its nearest neighbour is found and the
difference in their velocities used to estimate the relative velocity of collisions.
%%% Keplerian shear has been accounted for, i.e., by first multiplying $v_2$ by 
%%% $v_{\rm{k}}(r_1)/v_{\rm{k}}(r_2)$; though this accounts for r2/r1 it doesn't
%%% account for theta2-theta1 or phi2-phi1 for which the velocities need to be
%%% calculated in the frame rotating with Keplerian velocity}).
Taking relative velocities to the 8/3 power and averaging for each annulus results in
$\langle v_{\rm{rel}}^{8/3} \rangle (r)$.
The mean collision rate for individual planetesimals of size $D_{\rm{max}}$
in this population is then calculated using
\begin{equation}
  R_{\rm{cc}}(D_{\rm{max}}) = K M_{\rm{tot}} \sum_r
                              \bar{\sigma}(r) \bar{\sigma}_{\rm{v}}(r)
                              \langle v_{\rm{rel}}^{8/3} \rangle (r),
  \label{eq:rccfin}
\end{equation}
where $\sum_{r}$, and similar notation elsewhere, means to sum over radius,
and the extra $\bar{\sigma}(r)$ term (compared with equation~\ref{eq:rccdmax2})
accounts for the fraction of time the planetesimal spends in different annuli.

One further simplification is possible to this collision rate by expressing this in
terms of $\bar{r}=r/a$ using the fact that $\bar{\sigma}(r)=\bar{\sigma}(\bar{r})$,
and that relative velocity is proportional to the Keplerian velocity at $r=a$ times some
function of $e$ and $\bar{r}$ (see equation~\ref{eq:vrelanal}) so that
\begin{eqnarray}
  R_{\rm{cc}}(D_{\rm{max}}) & = & K M_{\rm{tot}} v_{\rm{k}}^{8/3}(a) a^{-3} (4\pi 
                                  I_{\rm{max}})^{-1} S(e), \label{eq:rccs} \\
  S_{\rm{2d}}(e) & = & \sum_{\bar{r}} 
                          \bar{r}^{-2} d\bar{r}^{-1}
                          \bar{\sigma}(\bar{r})^2
                          \langle [v_{\rm{rel}}/v_{\rm{k}}(a)]^{8/3} \rangle (\bar{r}), 
                          \label{eq:s2d}
\end{eqnarray}
where $S(e)=S_{\rm{2d}}(e)$ in equation (\ref{eq:rccs}) in the 2D approximation.

\begin{figure}
  \begin{center}
    \vspace{-0.2in}
    \begin{tabular}{c}
      \hspace{-0.5in} \psfig{figure=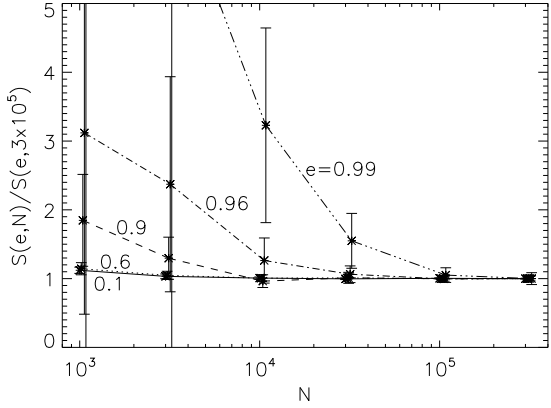,height=2.4in}
    \end{tabular}
    \caption{The factor $S_{\rm{2d}}(e)$ from equation~(\ref{eq:s2d})
    determined numerically using different numbers of planetesimals $N$ for $e=$0.1, 0.6,
    0.9, 0.96 and 0.99, compared with that determined using $N=3\times 10^5$.
    For each value of $N$, the calculation was repeated 10 times 
    to determine the $1\sigma$ uncertainty in the value derived in this way, and this is shown 
    by the error bars.
    For all eccentricities below 0.99 the calculation has converged for
    $N > 3 \times 10^5$.
    The number of radial bins was set at 40 for this calculation, which similar plots
    show is sufficient for this eccentricity range.}
   \label{fig:sevsnpp}
  \end{center}
\end{figure}

Successful implementation of this routine requires that the number of planetesimals ($N$) is
sufficiently large for the relative velocity of encounters to be well approximated by the
difference in the velocities with nearest neighbours.
The number of annuli ($N_{\rm{ann, r}}$) must also be sufficiently large to resolve the
radial variations in collision rate between pericentre and apocentre.
To assess this, Fig.~\ref{fig:sevsnpp} plots $S(e)$ as a function of
$N$ for $N_{\rm{ann, r}}=40$, normalised to the value expected when $S(e)$ is calculated
with $N=3\times10^5$ and $N_{\rm{ann, r}}=40$.
It is evident that the solution converges for $N>10^5$ for $e \leq 0.99$.
A similar plot showing the effect of changing the number of radial bins shows $N_{\rm{ann, r}}=40$
is sufficient for this eccentricity range, with values as low as 10 also giving results within
5\%.

%%%%%%%%%%%%%%%%%%%%%%%%%%%%%%%%%%%%%%%%%%%%%%%%%%%%%%
\subsubsection{Analytical collision rates}
\label{sss:analytical}
The 2D collision rates can also be derived analytically for a population with a common
semimajor axis and eccentricity, as the various factors
in equation (\ref{eq:s2d}) are the consequence of 2 body Keplerian motion. 
The distribution of cross-sectional area is determined by the rate of radial
motion $\bar{\sigma}(\bar{r})/d\bar{r} = 2a/(t_{\rm{per}}dr/dt)$, where $t_{\rm{per}}$
is the orbital period, giving:
\begin{equation}
  \bar{\sigma}/(d\bar{u}) = - \pi^{-1} \bar{u}^{-2} [\bar{u}^2(e^2-1)+2\bar{u}-1]^{-0.5},
  \label{eq:sigbaranal}
\end{equation}
where $\bar{u}=a/r$.
The relative velocity of collisions at each radius has a bimodal distribution with half of 
collisions occurring at zero velocity (for planetesimals moving in the same direction), and the
remainder at a velocity given by $\sqrt{2-2\cos{\phi}}$ times the orbital velocity at that
radius, where $\phi$ is the angle between the velocity vectors for planetesimals moving in 
different directions (e.g., those returning to pericentre colliding with those that have just 
passed pericentre), giving:
\begin{equation}
  v_{\rm{rel}}/v_{\rm{k}}(a) = 2 [ \bar{u}^2(e^2-1)+2\bar{u}-1 ]^{0.5},
  \label{eq:vrelanal}
\end{equation}
noting that $\langle [v_{\rm{rel}}/v_{\rm{k}}(a)]^{8/3} \rangle (\bar{r})$ is 
$0.5$ times the expression above to the $8/3$ power. 
Putting these expressions into equation (\ref{eq:s2d}) and integrating over the range
$\bar{u} = (1-e)^{-1}$ to $(1+e)^{-1}$ gives the relevant factor in the equation for
the collision rate as:
\begin{equation}
  S_{\rm{2d}} = 0.54 e^{5/3} (1-e^2)^{-4/3}.
  \label{eq:s2danal}
\end{equation}

%%%%%%%%%%%%%%%%%%%%%%%%%%%%%%%%%%%%%%%%%%%%%%%%%%%%%%
\subsubsection{Where do collisions occur, and at what velocity?}
\label{sss:wherecoll}

\begin{figure}
  \begin{center}
    \vspace{-0.2in}
    \begin{tabular}{cc}
      \textbf{(a)} & \hspace{-0.5in} \psfig{figure=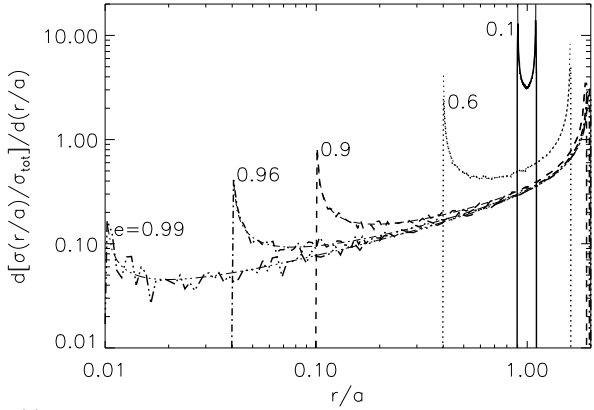,height=2.4in} \\
      \textbf{(b)} & \hspace{-0.5in} \psfig{figure=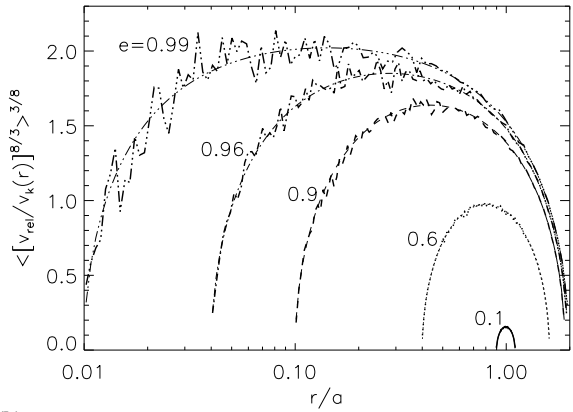,height=2.4in} \\
      \textbf{(c)} & \hspace{-0.5in} \psfig{figure=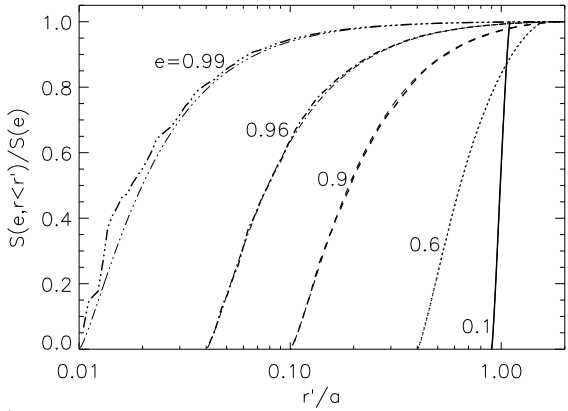,height=2.4in}
    \end{tabular}
    \caption{Radial distribution of collision rate for mutual collisions amongst a population
    of planetesimals with common pericentre and apocentre distances.
    The plots indicate the contributions of different factors in the collision rate equation
    (\ref{eq:s2d}) as a function of radius for a population with eccentricities of
    0.1, 0.6, 0.9, 0.96, 0.99.
    The results of both the Monte-Carlo simulation and the analytical calculation are
    plotted showing excellent agreement.
    \textbf{(a)} The distribution of cross-sectional area density,
    $\bar{\sigma}(\bar{r})/d\bar{r}$;
    the integral under the curve for each population is 1.
    \textbf{(b)} The average collision velocities of planetesimals at different
    radii relative to the local Keplerian velocity
    $[\langle [v_{\rm{rel}}/v_{\rm{k}}(r)]^{8/3} \rangle (\bar{r})]^{3/8}$.
    \textbf{(c)} The fraction of $S_{\rm{2d}}(e)$ that comes from radii below
    $r'$.}
   \label{fig:sevsr}
  \end{center}
\end{figure}

The first question we can answer with this model is where most of the collisions
occur.
This can be worked out from the distributions plotted in Figure~(\ref{fig:sevsr}).
Although there is also a small density enhancement at pericentre, 
planetesimals spend the majority of their time at apocentre
(Figure~\ref{fig:sevsr}a; see, e.g., Fig.~4b of Sykes 1990).
Nevertheless, the apocentric contribution to the overall collision rate
is diminished due to both the $r^{-2}$ term in equation~(\ref{eq:s2d}) and the
higher relative velocities at lower radii
(Figure~\ref{fig:sevsr}b), where it should be noted that
the factor in equation~(\ref{eq:s2d}) is relative to the Keplerian
velocity at $r=a$ and so is Figure~\ref{fig:sevsr}b multiplied by $\sqrt{a/r}$
then to the $8/3$ power.
The net result is that the majority of the collisions occur close to pericentre 
for all except the lowest eccentricities (Figure~\ref{fig:sevsr}c).
The Monte-Carlo simulation provides results in excellent agreement with the
analytical calculation (\S \ref{sss:analytical}) from which it can be found that 90\% 
of collisions occur at $r/a<(1-e^2)/(1-0.72e)$.

As noted in \S \ref{sss:analytical}, the distribution of collision velocities in this 
population is not uniform.
Except for planetesimals that are close to pericentre or apocentre, relative velocities
in each annulus have a bimodal distribution, as for planetesimals that have recently passed 
pericentre there is a low relative velocity population that have also recently passed 
pericentre and a high relative velocity population that are returning from apocentre to 
pericentre (as noted in \S \ref{sss:analytical}).
Thus the relative velocity of the high relative velocity population is approximately
$2^{3/8}$ times that shown in Figure~\ref{fig:sevsr}b, and remains within the range
1-2.5 times the local Keplerian velocity between apocentre and pericentre for $e=0.6-0.99$.
This means that, for a population with pericentre at 1AU, collisions occur at velocities
of $10-100$ km s$^{-1}$, and that relatively small planetesimals can cause
destructive collisions (equation~\ref{eq:xc}).

%%%%%%%%%%%%%%%%%%%%%%%%%%%%%%%%%%%%%%%%%%%%%%%%%%%%%%
\subsubsection{Collision rate and remaining mass vs eccentricity}
\label{sss:sevse}

\begin{figure}
  \begin{center}
    \vspace{-0.2in}
    \begin{tabular}{cc}
      \textbf{(a)} & \hspace{-0.5in} \psfig{figure=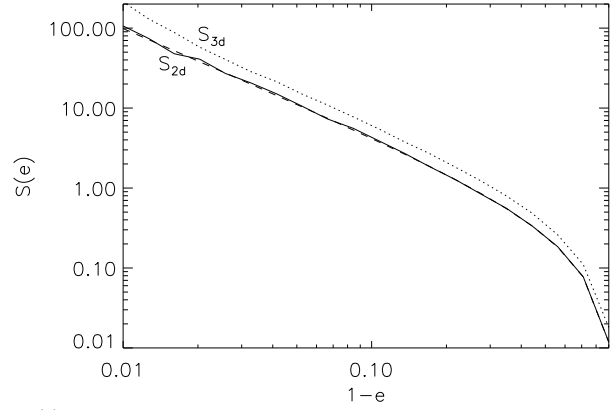,height=2.4in} \\
      \textbf{(b)} & \hspace{-0.5in} \psfig{figure=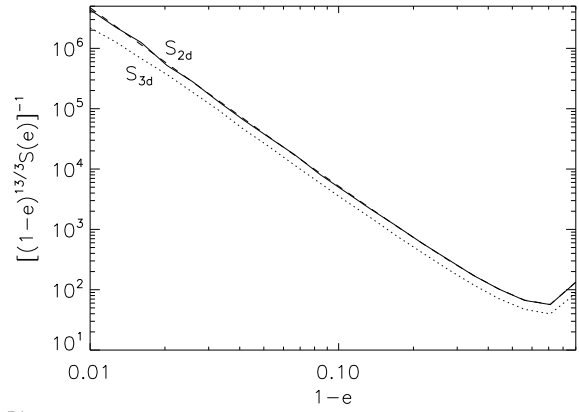,height=2.4in}
    \end{tabular}
    \caption{Dependence of collision lifetime on eccentricity.
    For a disk in which mass and semimajor axis are fixed collision lifetime
    $\propto 1/S(e)$.
    The function $S(e)$ is shown in \textbf{(a)}, both calculated in the 2D 
    approximation ($S_{\rm{2d}}(e)$, solid line), and with the full 3D calculation 
    ($S_{\rm{3d}}(e)$, dotted line), as well as the analytical calculation of the
    2D approximation (dashed line which lies under the solid line).
    For a disk in which mass and pericentre distance are fixed collision lifetimes vary
    $\propto [(1-e)^{13/3}S(e)]^{-1}$.
    This function is shown in \textbf{(b)} with the same origin for the 
    different linestyles as \textbf{(a)}.}
   \label{fig:sevse}
  \end{center}
\end{figure}

Figure~\ref{fig:sevse}a shows how $S_{\rm{2d}}(e)$ varies with eccentricity, where the
results of the Monte-Carlo simulation closely follow the predictions of the analytical
calculation in equation~(\ref{eq:s2danal}).
The factor $S(e)$ can be readily used to assess how changing the eccentricity of a 
planetesimal population affects its collisional lifetime using equation~(\ref{eq:rccs}).
It can also be used to consider how eccentricity affects the amount of disk 
mass that can remain at late times.
For the assumptions about the evolution of the size distribution discussed in
\S \ref{ss:evolsd}, the evolution of planetesimal belt mass from its initial value of
$M_{\rm{tot0}}$ can be calculated from the collisional rate using the equation
$dM_{\rm{tot}}/dt = -M_{\rm{tot}}R_{\rm{cc}}(D_{\rm{max}})$, which can
be solved to give the mass at time $t$ in years to be
\begin{equation}
  M_{\rm{tot}}/M_{\rm{tot0}} = [ 1 + 
    ( R_{\rm{cc}}(D_{\rm{max}}) M_{\rm{tot0}} / M_{\rm{tot}} ) t ]^{-1}.
\end{equation}
This means that at late times the remaining mass converges to a value of
(Wyatt et al. 2007a):
\begin{eqnarray}
  M_{\rm{late}} & = & t^{-1}[M_{\rm{tot}}/R_{\rm{cc}}(D_{\rm{max}})], \\
                & = & (4\pi I_{\rm{max}} / K) [a^3 / v_{\rm{k}}^{8/3}(a)] [S(e) t]^{-1}, 
   \label{eq:mlate1}
\end{eqnarray}
where late means $t \gg 1/R_{\rm{cc}}(D_{\rm{max}})$.

Here we illustrate this in two ways.
First we consider a disk in which eccentricity is varied, but the semimajor axis and disk mass 
are kept constant.
The collisional lifetime of such a disk is $\propto 1/S(e)$, meaning that increasing
eccentricity results in a shorter collisional lifetime.
The mass remaining at late times is also $\propto 1/S(e)$, which from Fig.~\ref{fig:sevse}
decreases rapidly with eccentricity, because of the increased collision rate of material at a 
pericentre which tends to smaller radii.

However, if pericentre location and disk mass are kept fixed as eccentricity is
increased, collisional lifetime instead varies $\propto [(1-e)^{13/3}S(e)]^{-1}$.
Rewriting equation~\ref{eq:mlate1} gives
\begin{equation}
  M_{\rm{late}} = [1.47 \times 10^{-11} I_{\rm{max}} / K]
    M_\star^{-4/3} q^{13/3} t^{-1} [(1-e)^{13/3}S(e)]^{-1}, \label{eq:mlate2}
\end{equation}
so that mass remaining is also $\propto [(1-e)^{13/3}S(e)]^{-1}$, and this function is
plotted in Fig.~\ref{fig:sevse}b.
Thus both collision lifetime and mass remaining increases as eccentricity is
increased $\propto e^{-5/3}(1+e)^{4/3}(1-e)^{-3}$.
Primarily this is because planetesimals spend a larger fraction of their orbit at
large radii resulting in both a longer time between collisions, and a lower density
of colliders in the region where collisions occur near pericentre;
as illustrated in Fig.~\ref{fig:sevsr}, increasing eccentricities also results
in an enhanced collision velocity at pericentre which offsets this to some
extent, but there is no significant change in collision velocities once
eccentricities have increased beyond 0.4.

%%%%%%%%%%%%%%%%%%%%%%%%%%%%%%%%%%%%%%%%%%%%%%%%%%%%%%
\subsubsection{Comparison with low eccentricity approximation}
\label{sss:complowe}
The collision rate in the low eccentricity approximation (\S \ref{ss:lowe})
can also be expressed in the form of equation~\ref{eq:rccs}, where $a=r$ and
$S(e)=S_{\rm{lowe}}(e)$:
\begin{equation}
  S_{\rm{lowe}} = 3.0 (dr/r)^{-1} e^{8/3}.
\end{equation}
This could be plotted on Fig.~\ref{fig:sevse}, e.g., assuming that the width
of the ring is $dr/r=2e$, which would show agreement within an order of magnitude
for $e=0.1-0.98$.
However, this comparison is less instructive than noting that the
masses (and fractional luminosities) remaining at late times in Wyatt et al.
(2007a) were derived using $e=0.05$ and $dr/r=0.5$ for which
$S_{\rm{lowe}}=2 \times 10^{-3}$ and $[(1-e)^{13/3}S_{\rm{lowe}}(e)]^{-1}=610$.
Thus to increase the amount of mass that can remain at late times above the
values in that paper, assuming the radius inferred for the population corresponds to
the location of the pericentre of an eccentric ring, would require eccentricities higher
than 0.78, with a factor of $>100$ increase for $e>0.96$.
This illustrates the point that collisional lifetimes (and remaining mass) can be
increased both by increasing eccentricity and by spreading material over a broader range
of radii.

%%%%%%%%%%%%%%%%%%%%%%%%%%%%%%%%%%%%%%%%%%%%%%%%%%%%%%
\subsubsection{3D calculation}
\label{sss:3dapprox}
For the 3-dimensional calculation, it is further assumed that the longitudes of
ascending node $\Omega$ are random and that inclinations are randomly chosen
from the range 0 to $I_{\rm{max}}$, and these were used to calculate 3-dimensional
positions and velocities for the planetesimals.
Since collision rate is then also a function of latitude, the disk was further split
in latitude into $N_{\rm{ann, \phi}}$ bins at $\phi \pm d\phi/2$.
The planetesimal positions were used to derive $\bar{\sigma}(r,\phi)$, which is the
fraction of the total cross-sectional area in the annulus at $r \pm dr/2$
and $\phi \pm d\phi/2$.
Dividing this by the volume of the annulus, $dV_{\rm{3d}} = 2\pi r^3 (dr/r) \cos{\phi} d\phi$, 
gives the normalised volume density $\bar{\sigma}_{\rm{v}}(r,\phi)$.
The mean velocities are also a function of $\phi$, $\langle v_{\rm{rel}}^{8/3} \rangle (r,\phi)$).
This results in a collision rate given by equation (\ref{eq:rccs}) in which
$S(e)=S_{\rm{3d}}(e,I_{\rm{max}})$, where
\begin{eqnarray}
  S_{\rm{3d}}(e,I_{\rm{max}}) & = & \sum_{\bar{r}}
                          \bar{r}^{-2} d\bar{r}^{-1}
                       \sum_{\phi}
                          (2I_{\rm{max}}/\cos{\phi} d\phi)
\times \nonumber \\
   & &
                          \bar{\sigma}(\bar{r},\phi)^2
          \langle [v_{\rm{rel}}/v_{\rm{k}}(a)]^{8/3} \rangle 
                (\bar{r},\phi). \label{eq:s3d}
  \label{eq:rccfin2}
\end{eqnarray}

We can now answer how important it is to account for the
3-dimensional structure of the disk.
For $I_{\rm{max}}=0.05$ and using $N_{\rm{ann, \phi}}=9$ it was found that
$S_{\rm{3d}}(e,I_{\rm{max}})/S_{\rm{2d}}(e)$ has a constant value of
$1.43 \pm 0.04$ for eccentricities in the range $0.1-0.98$
(see Figure~\ref{fig:sevse}).
We attribute this offset to the fact that an even distribution of
inclinations between 0 and $I_{\rm{max}}$ results in a higher density in the mid-plane
(to achieve a uniform density we could have used the inclination distribution function
given by equation 2.20 of Krivov et al. 2005).
Thus the 2-dimensional results give a good approximation of the collision
rates in a 3-dimensional disk, as long as the parameter $I_{\rm{max}}$ used
in this calculation is interpreted as there being a uniform distribution
of inclinations between 0 and $1.43I_{\rm{max}}$.
However, for low eccentricities $e \ll I_{\rm{max}}$, where collision velocities are
not due only to radial and azimuthal motion, but also latitudinal motion, 
a full 3-dimensional calculation would be necessary.

%%%%%%%%%%%%%%%%%%%%%%%%%%%%%%%%%%%%%%%%%%%%%%%%%%%%%%
%%%%%%%%%%%%%%%%%%%%%%%%%%%%%%%%%%%%%%%%%%%%%%%%%%%%%%
\section{Evolution scheme for eccentric planetesimal swarm}
\label{s:4}
The single pericentre and apocentre distance population of \S \ref{s:2} was
necessarily an idealised case given that the eccentric populations in
the Solar System have a range of eccentricities.
This section describes how the modelling method of \S \ref{s:2} can be
generalised for such a situation, both to calculate the collision rates between
populations with different eccentricities and semimajor axes (\S \ref{ss:general}),
and to use these rates to model the collisional evolution of a planetesimal belt
with a distribution of orbital elements (\S \ref{ss:implem}).
The approach to modelling the collisional evolution is similar to the
kinetic model employed by Krivov et al. (2005) in that we consider the evolution
of the phase space distribution, which here is just two dimensional and
defined by $M_{\rm{tot}}(q,Q)$.

%%%%%%%%%%%%%%%%%%%%%%%%%%%%%%%%%%%%%%%%%%%%%%%%%%%%%%
\subsection{Collisions between two single pericentre-apocentre populations}
\label{ss:general}
The parameter space is divided into
cells of size $q \pm dq/2$ and $Q \pm dQ/2$, hereafter simply referred to
as cell $(q,Q)$, where $M_{\rm{tot}}(q,Q)$ is the total mass
in that cell.
Cells are logarithmically spaced.
For now it is assumed that the distribution of inclinations is the same
for each cell, and other angles are randomised as before, and so
where needed the spatial distribution from material in each
cell ($\bar{\sigma}(r)(q,Q)$ and $\bar{\sigma}_{\rm{v}}(r)(q,Q)$) can
be ascertained in one of the ways described in \S \ref{ss:highe}.

The scheme described in \S \ref{s:2} can be used in a similar manner to work
out the rate of catastrophic impacts onto planetesimals of size $D_{\rm{max}}$
in cell $(q_1,Q_1)$ from planetesimals in cell $(q_2,Q_2)$
\begin{equation}
  R_{\rm{cc}}(D_{\rm{max}},q_1,Q_1;q_2,Q_2) =
       M_{\rm{tot}}(q_2,Q_2)
       \bar{R}_{\rm{cc}}(D_{\rm{max}},q_1,Q_1;q_2,Q_2).
  \label{eq:rcol12}
\end{equation}
For the 2-dimensional approximation the normalised collision rate is
\begin{eqnarray}
  \bar{R}_{\rm{cc}}(D_{\rm{max}},q_1,Q_1;q_2,Q_2) & = &
    K \sum_r
     \bar{\sigma}(r)(q_1,Q_1)
   \times \nonumber \\ & & 
     \bar{\sigma}_{\rm{v}}(r)(q_2,Q_2)
    \langle v_{\rm{rel[1,2]}}^{8/3} \rangle (r),
  \label{eq:rcolbar12}
\end{eqnarray}
where $\langle v_{\rm{rel[1,2]}}^{8/3} \rangle (r)$ is the mean 
of the relative velocities to the $8/3$ power between planetesimals in
the different cells at that radius (which can, e.g., be worked out
using the Monte-Carlo method of \S \ref{ss:highe}).

The computation of $\bar{R}_{\rm{cc}}(D_{\rm{max}},q_1,Q_1;q_2,Q_2)$
can be made more efficient noting that 
$\bar{R}_{\rm{cc}}(D_{\rm{max}},q_1,Q_1;q_2,Q_2) =
\bar{R}_{\rm{cc}}(D_{\rm{max}},q_2,Q_2;q_1,Q_1)$ and 
that some combinations of $q_1,Q_1,q_2,Q_2$ are either unphysical
or non-overlapping, as well as by only including in the Monte Carlo
calculation planetesimals where the orbits from the two cells overlap.
A similar simplification to that used to derive equations
(\ref{eq:rccs})-(\ref{eq:s2d})
can also be employed so that collision rate is given by
\begin{eqnarray}
  \bar{R}_{\rm{cc}}(D_{\rm{max}},q_1,Q_1;q_2,Q_2) & = &
      (K/2.1\times 10^{-14}) M_\star^{4/3} q_1^{-13/3}
      \times \nonumber \\ & &
      I_{\rm{max}}^{-1}
      T(e_1,e_2,q_2/q_1).
      \label{eq:rccs2}
\end{eqnarray}
Here-on we plot the function $T(e_1,e_2,q_2/q_1)$ defined in this way
rather than $\bar{R}_{\rm{cc}}$, noting that the situation
described in \S \ref{ss:highe} corresponds to $T(e,e,1)$ which
must therefore be equal to
$1.43 \times 10^{-3} (1-e)^{13/3} S_{\rm{2d}}(e)$.

%%%%%%%%%%%%%%%%%%%%%%%%%%%%%%%%%%%%%%%%%%%%%%%%%%%%%%
\subsection{Comparison with previous collision rate calculations}
\label{ss:comp}
The collision rates described in \S \ref{ss:general} 
use the 2-dimensional approximation, but can be readily modified
for a 3-dimensional calculation in a manner similar to \S \ref{sss:3dapprox}.
It is also possible to modify the assumptions to choose an inclination
distribution that spans a narrow range $I_{\rm{min}}$ to $I_{\rm{max}}$.
In this case the collision rates we derive should be comparable with those
of other authors who consider interactions between planetesimals on two 
orbits each of which is defined by a semimajor axis, eccentricity and
inclination (with other angles assumed to be randomly distributed).

\begin{table}
%\begin{minipage}{160mm}
  \begin{center}
    \caption{Intrinsic collision probabilities in $10^{-18}$ km$^{-2}$ yr$^{-1}$
    between the Astrid and objects from Table II of
    Dell'Oro \& Paolicchi (1998) for comparison of our results with those
    of that paper and with those of Bottke \& Greenberg (1993).}
    \begin{tabular}{lccc}
        \hline
        Object        & BG93  & DP98  & This paper \\
        \hline
        1948 EA       & 3.16  & 3.19  & 3.17   \\
        Apollo        & 3.58  & 3.58  & 3.58   \\
        Adonis        & 4.51  & 4.52  & 4.35   \\
        1950 DA       & 3.69  & 3.76  & 3.76   \\
        Encke         & 3.36  & 3.42  & 3.43   \\
        Brorsen       & 0.93  & 0.94  & 0.94   \\
        Grigg-Mellish & 0.022 & 0.022 & 0.022  \\
        Temple-Tuttle & 0.62  & 0.62  & 0.60   \\
        Neujmin       & 0.94  & 0.93  & 0.93   \\
        Schaumasse    & 1.13  & 1.15  & 1.15   \\
        Pons Brooks   & 0.041 & 0.041 & 0.041  \\
        \hline
    \end{tabular}
    \label{tab:tab1}
  \end{center}
%\end{minipage}
\end{table}

To carry out such a comparison we also need to modify the calculation to
derive the more typically quoted quantity of the intrinsic collision
probability, $P_i$, which is the probability of impact per unit time
divided by $\tau^2$ where $\tau=(D+D_{\rm{im}})/2$ (Wetherill 1967).
Our method included the $\tau^2$ factor from the outset
(see equation~\ref{eq:f}) and integrated the collision rate over
the size distribution capable of causing catastrophic impacts, whereas
this was accounted for at a later stage by other methods. 
Using our notation, intrinsic probability in $10^{-18}$ km$^{-2}$ yr$^{-1}$
is given by 
\begin{equation}
  P_i = 0.030 \sum_{r,\phi} 
     \bar{\sigma}(r,\phi)(q_1,Q_1)
     \bar{\sigma}_{\rm{v}}(r,\phi)(q_2,Q_2)
    \langle v_{\rm{rel[1,2]}} \rangle (r,\phi),
  \label{eq:pi}
\end{equation}
where the constant includes both a conversion between units and an
extra factor $\pi$ (since our calculation considers cross-sectional
area rather than $\tau^2$).
In Table~\ref{tab:tab1} we compare the intrinsic collision probabilities
we find using the Monte-Carlo approach with 200,000 planetesimals in each
population with those of Dell'Oro \& Paolicchi (1998) and Bottke \& Greenberg (1993)
showing excellent agreement between the methods.

\begin{figure}
  \begin{center}
    \vspace{-0.2in}
    \begin{tabular}{c}
      \psfig{figure=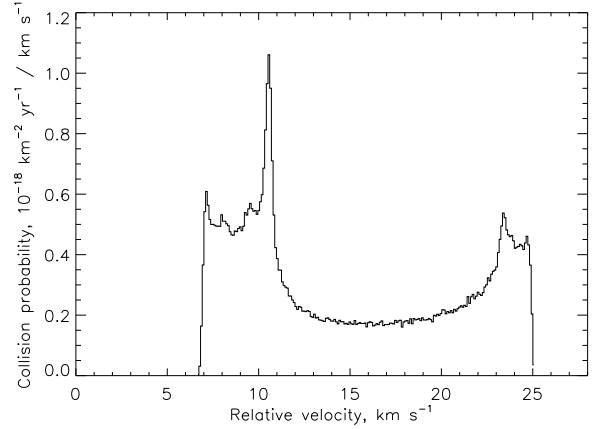,height=2.4in}
    \end{tabular}
    \caption{Contribution of different relative velocities to the
    intrinsic collision probability for two planetesimals, one with
    orbital elements $a=3.42$AU, $e=0.578$, $I=0.435$rad and the other
    with $a=1.59$AU, $e=0.056$, $I=0.466$rad for comparison with
    Fig.~1 of Bottke et al. (1994) and Fig.~1 of Dell'Oro \& Paolicchi (1998).}
   \label{fig:bottke94}
  \end{center}
\end{figure}

As noted by Bottke et al. (1994) (their point I.2), to utilise intrinsic collision
probabilities requires knowledge of the velocity probability distribution
and it is not sufficient to assume a single mean relative velocity of collisions.
This is automatically included in our method, since we keep track of the
velocities encountered by a planetesimal at each location and these are
used to work out the amount of cross-sectional area of impactors at that location
that are able to cause a catastrophic collision. 
To illustrate that our Monte-Carlo method computes an accurate velocity probability
distribution, here we reproduce Figure 1 of Bottke et al. (1994) (see also
Fig.~1 of Dell'Oro \& Paolicchi 1998) in our
Figure~\ref{fig:bottke94}.

Similarly we find that we can reproduce Figs~5 and 6 of Krivov et al. (2005)
(not shown here) noting that their $\Delta$ and $\bar{v}_{\rm{imp}}$ are given by
\begin{eqnarray}
  \Delta(a_1,e_1,a_2,e_2) & = & \sum_{r,\phi} \bar{\sigma}(r,\phi)(q_1,Q_1)
    \bar{\sigma}_{\rm{v}}(r,\phi)(q_2,Q_2), \\
  \bar{v}_{\rm{imp}} & = & \sum_{r,\phi} \bar{\sigma}(r,\phi)(q_1,Q_1)
    \bar{\sigma}_{\rm{v}}(r,\phi)(q_2,Q_2)
  \times \nonumber \\ & & 
    \langle v_{\rm{rel[1,2]}} \rangle (r,\phi) / \Delta(a_1,e_1,a_2,e_2),
\end{eqnarray}
in our notation, although we should point out again that such a mean collision
velocity must be used with care as a range of velocities contribute to
the collision rate (see Fig.~\ref{fig:bottke94}).

In summary we conclude that, for the same assumptions, our method for
calculating collision rates produces results that are in agreement with the
most accurate methods available in the literature, and that although this
method was derived with the intention of studying high eccentricity orbits,
it is also applicable regardless of the magnitude of the eccentricity.

%%%%%%%%%%%%%%%%%%%%%%%%%%%%%%%%%%%%%%%%%%%%%%%%%%%%%%
\subsection{Implementation of evolution}
\label{ss:implem}
The factor $\bar{R}_{\rm{cc}}(D_{\rm{max}},q_1,Q_1;q_2,Q_2)$ (equation 
\ref{eq:rcolbar12}) depends only on the way in which the parameter space
is divided up.
Thus an approach was implemented in which this factor was worked out ahead of time,
and then used to work out the catastrophic collision rate for the largest
planetesimals in cell $(q,Q)$ from planetesimals in all other cells:
\begin{equation}
  R_{\rm{cc}}(D_{\rm{max}},q,Q) = 
  \sum_{q_2}
  \sum_{Q_2}
    M_{\rm{tot}}(q_2,Q_2)
    \bar{R}_{\rm{cc}}(D_{\rm{max}},q,Q;q_2,Q_2).
\end{equation}

Using the assumptions about the evolution of the size distribution
described in \S \ref{ss:evolsd}, this collision rate can then be used to
work out the mass which would be removed in a timestep $dt$, and the mass
distribution stepped forward in time.
\begin{equation}
  M_{\rm{tot}}(q,Q,t+dt) = M_{\rm{tot}}(q,Q,t)[1 - dt*R_{\rm{cc}}(D_{\rm{max}},q,Q)].
\end{equation}
The timestep $dt$ is chosen so that some fraction, say 0.1\%, of the mass is removed
from the most rapidly evolving cell in that timestep, i.e.,
\begin{equation}
  dt = 10^{-3}/\rm{max}[R_{\rm{cc}}(D_{\rm{max}},q,Q)].
\end{equation}

%%%%%%%%%%%%%%%%%%%%%%%%%%%%%%%%%%%%%%%%%%%%%%%%%%%%%%
\subsection{Application to scattered disk-like distribution}
\label{ss:sd}
One of the simplest examples of this evolution is its application
to a scattered disk-like distribution wherein planetesimals have
a common pericentre distance, $q$, but a range of apocentre distances.
Such a distribution might arise from planetesimals scattered by a planet
on a circular orbit at a radius just inside $q$.

\begin{figure}
  \begin{center}
    \vspace{-0.2in}
    \begin{tabular}{cc}
      \textbf{(a)} & \hspace{-0.5in} \psfig{figure=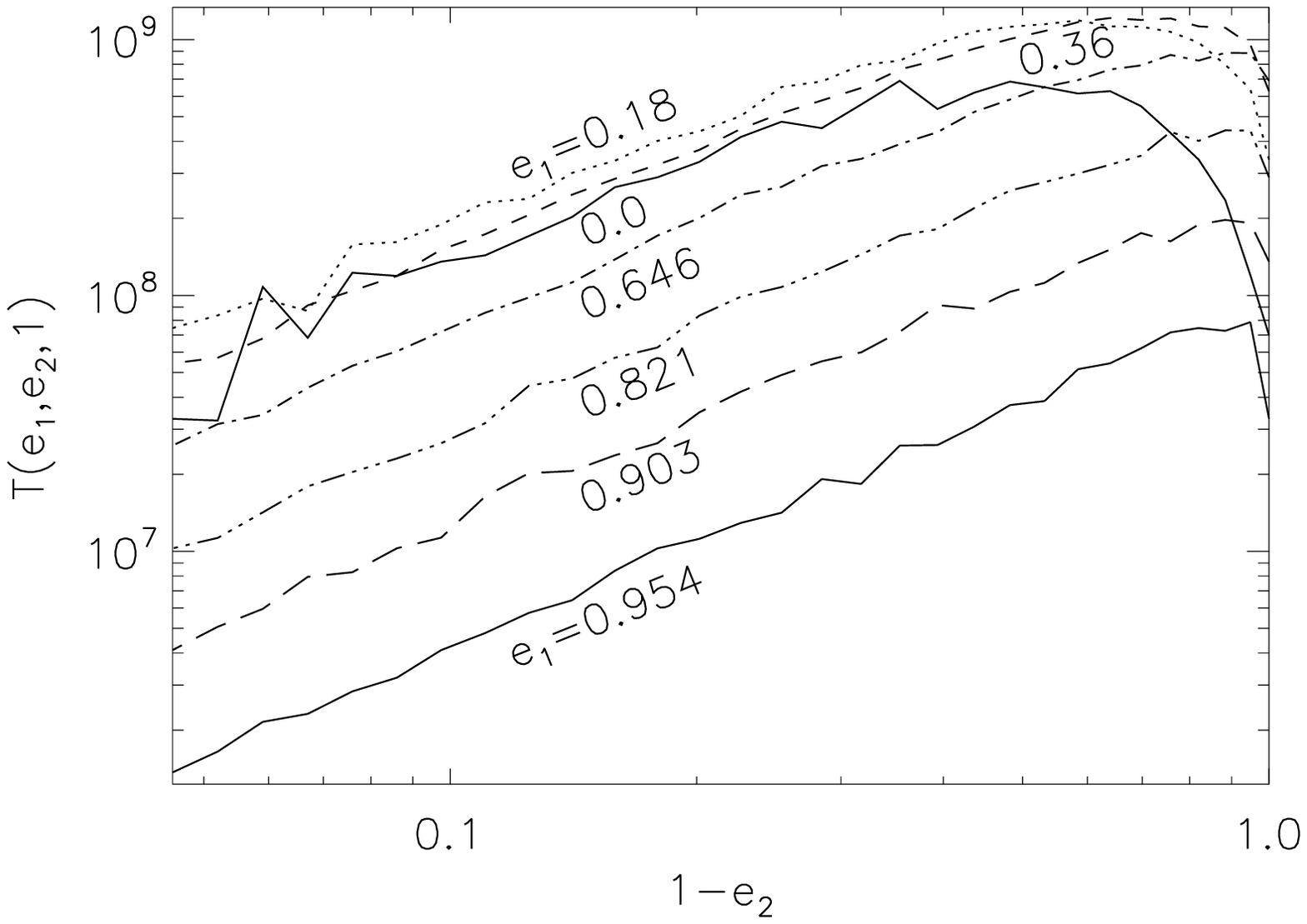,height=2.4in} \\
      \textbf{(b)} & \hspace{-0.5in} \psfig{figure=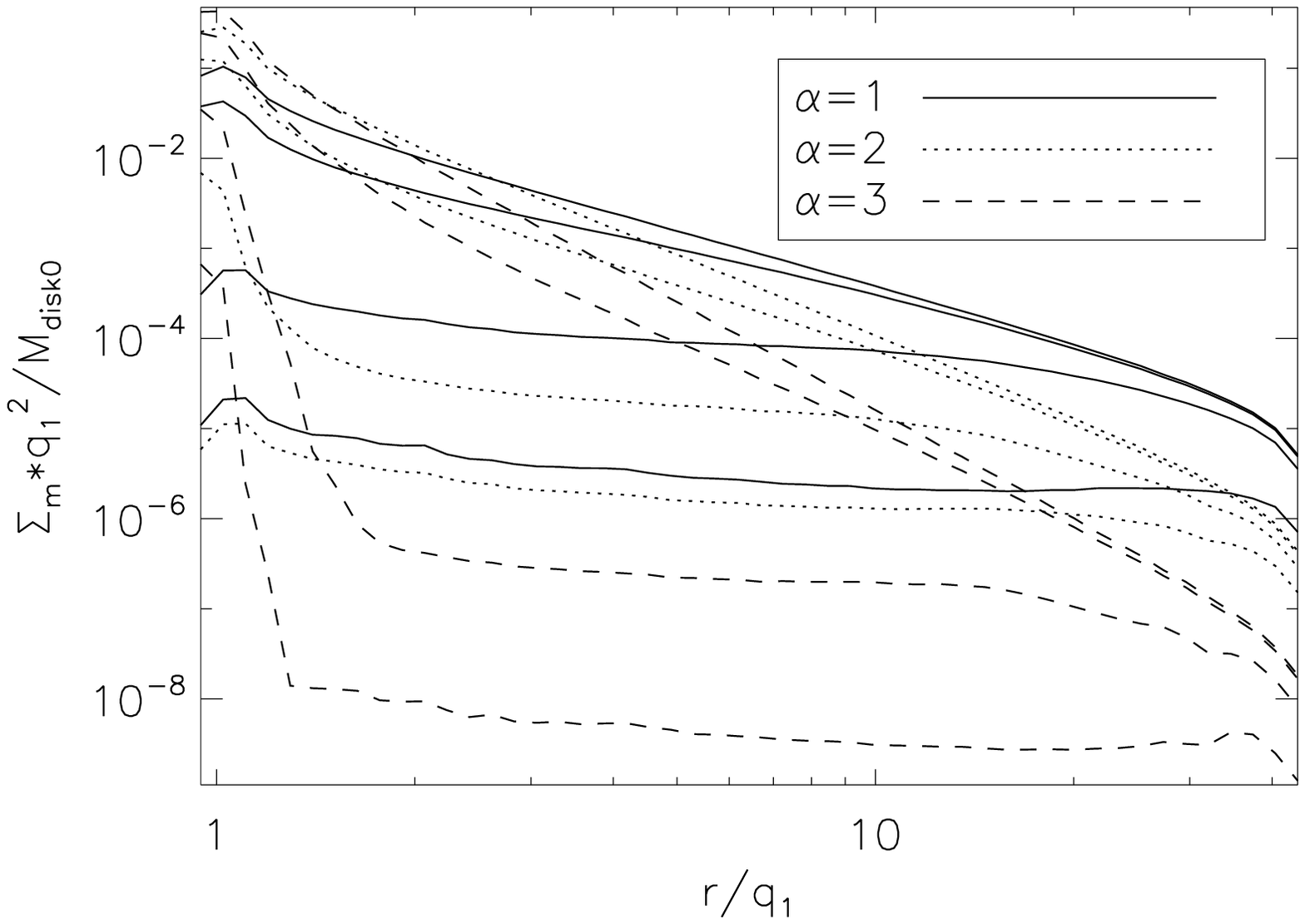,height=2.4in} \\
      \textbf{(c)} & \hspace{-0.5in} \psfig{figure=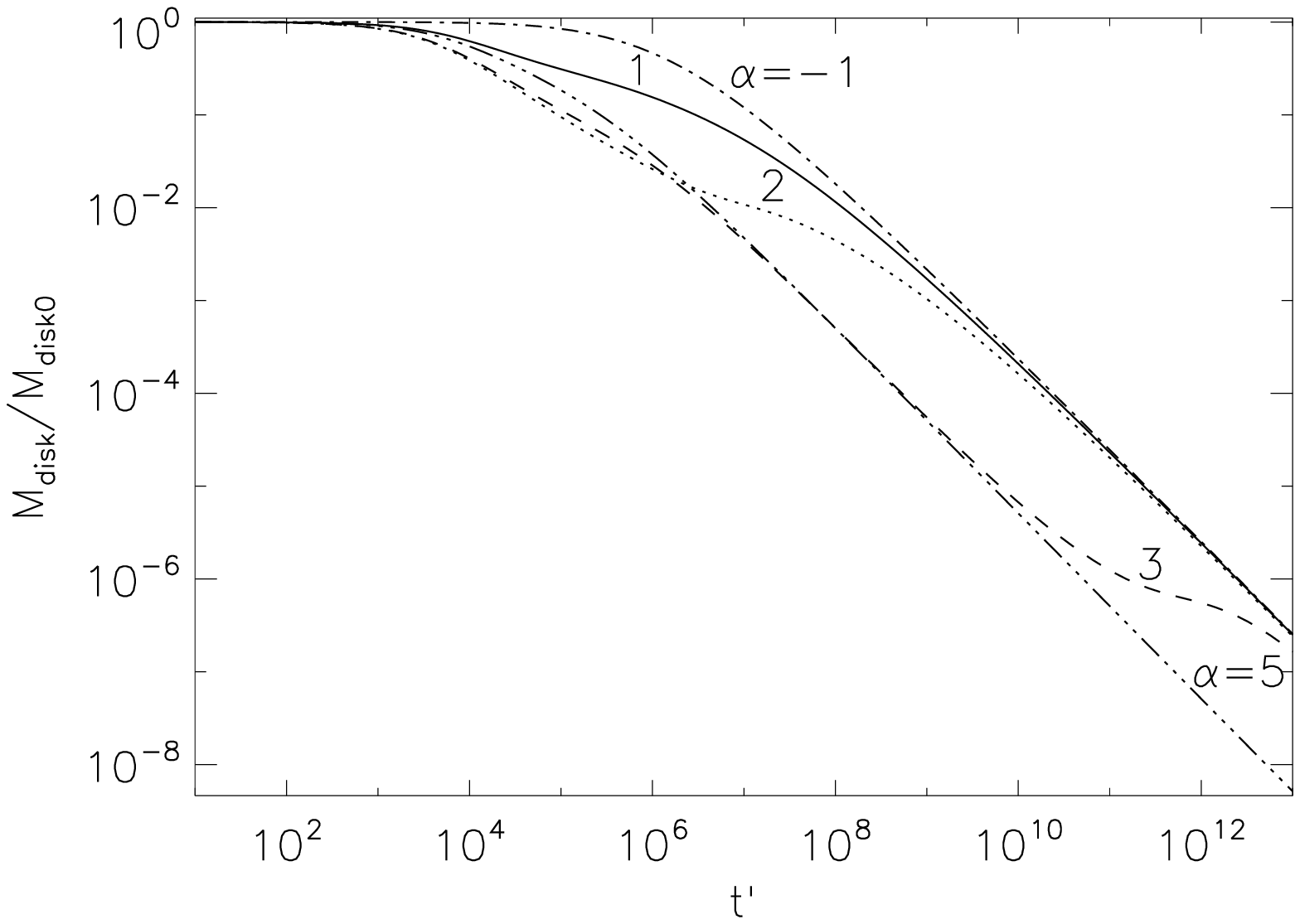,height=2.4in}
    \end{tabular}
    \caption{Collisional evolution of a scattered disk-like distribution.
    \textbf{(a)} The function $T(e_1,e_2,1)$ that through equation~(\ref{eq:rccs2})
    defines the rate of collisions
    for a planetesimal in population 1 interacting with planetesimals in population 2.
    \textbf{(b)} Evolution of the mass surface density distribution assuming an
    initial distribution of apocentres $n(Q) \propto Q^{-\alpha}$,
    where $\alpha=1, 2, 3$ is shown with solid, dotted and dashed lines respectively.
    The resulting evolution scales with $q_1$ and $M_{\rm{disk0}}$ as per the
    axis labels, and 4 lines are shown for each initial apocentre distribution,
    at times of $t'=0, 10^4, 10^6, 10^8$, where real time in
    years is related to $t'$ through equation~(\ref{eq:treal}).
    \textbf{(c)} Evolution of total disk mass for the distributions plotted in
    \textbf{(b)}, including also $\alpha=-1$ and 5.}
   \label{fig:sd}
  \end{center}
\end{figure}

Figure~\ref{fig:sd}a shows the function $T(e_1,e_2,1)$ for a population with
pericentres in the range $(1 \pm 0.1)q_1$ and apocentres in 30 logarithmically
spaced bins in the range $(1-50)q_1$.
This illustrates how for most planetesimals in the population, their collision rate
per unit mass of the colliding population is highest for collisions with the
lowest eccentricity population.
This is because the mass of a lower eccentricity colliding population is concentrated
in a smaller volume resulting in higher densities and so higher collision rates.
However, for the low eccentricity population it is noticeable that $T(0,e_2,1)$
peaks at eccentricities of around 0.5, and that the collision rate per unit mass
of the colliding population is lower with planetesimals on circular orbits.
This is because the higher density of the low eccentricity colliders is offset
by the lower relative velocities.
The reduction in relative velocities as eccentricity is decreased was already
noted in Figure~\ref{fig:sevsr}b, and its impact on collision rates evident in
Figure~\ref{fig:sevse}b.

To illustrate how such a population evolves through mutual collisions, 
we assume that the mass in orbits in the range $Q$ to $Q+dQ$ starts
off $\propto Q^{-\alpha}dQ$ with a total mass of $M_{\rm{disk0}}$ distributed
among apocentres in the range $Q/q=1-50$.
This assumption results in an initial distribution of mass surface density that falls off
approximately as $\Sigma_m \propto r^{-\alpha-1}$, because most of the mass is
concentrated at apocentre.
Figure~\ref{fig:sd}b shows the evolution of mass surface density for $\alpha=1,2,3$,
where equation~\ref{eq:rccs2} was used to normalise both radius and mass surface
density, meaning that the times plotted correspond to real times of
\begin{equation}
  t = (2.1 \times 10^{-14}/K) M_\star^{-4/3} (I_{\rm{max}}/0.05) q_1^{13/3} M_{\rm{disk}}^{-1} t'
  \label{eq:treal}
\end{equation}
years, where $M_{\rm{disk}}$ is in $M_\oplus$ and $q_1$ in AU.

The $\alpha=1$ evolution shows how the distribution tends to a flat distribution
in mass surface density.
This is because the low eccentricity population is rapidly depleted, with mass
becoming ever more concentrated in larger apocentre population that evolves
relatively slowly;
note from Fig.~\ref{fig:sevsr}a that a high eccentricity population would have
a fairly flat mass surface density distribution. 
The $\alpha=3$ evolution differs from that of $\alpha=1$ in that a bimodal population
is formed with a low eccentricity population causing the peak in surface density
at low $r$ and a high eccentricity population causing the flat surface density distribution.
Because the mass starts off concentrated in the low eccentricity population, its evolution
is unaffected by the material with large apocentres, and so it evolves due to mutual
collisions.
However, populations with higher eccentricities are rapidly depleted in collisions
with the low eccentricity population, with the highest eccentricities persisting the longest.
Eventually the low eccentricity population is depleted in mutual collisions so that
the mass is concentrated in the highest eccentricities even for $\alpha \geq 3$.

The evolution of total disk mass is shown in Fig.~\ref{fig:sd}c.
From this it can be seen that the evolution is slower for shallower apocentre distributions.
All distributions also tend to a mass evolution that falls off $\propto 1/t$,
meaning that the amount of mass remaining at late times is independent of both initial mass
and $\alpha$.
The amount of remaining mass lies between two values:
If the mass is in the lowest eccentricities (steep distributions at young ages) then that
mass is given by that expected from a low eccentricity population at $r=q$.
However, if the mass is in the highest eccentricities (all distributions at late ages) then that
mass is given by the mass expected for the highest eccentricities in the distribution;
since $e_{\rm{max}}=0.96$ in this simulation because $Q/q \leq 50$, this figure is in agreement with
the expectations of Fig.~\ref{fig:sevse}b and \ref{fig:sd}a.
Typically the evolution switches from having the mass in the lowest eccentricity
population to the highest eccentricity population.

Application of this can be readily seen for the case of the
scattered disk in the Kuiper belt.
For a pericentre at 30AU and an initial mass of $0.1M_\oplus$
we see that the evolutionary timescale is very long, since real time is
$25 \times 10^{6}t'$ (for $K=2.1 \times 10^{-14}$ and $I_{\rm{max}}=0.05$), so
that 4.5Gyr corresponds to $t'= 200$,
meaning that collisional mass loss would be expected to be very small over
the age of the Solar System, regardless of $\alpha$.
Had the scattered disk had 100 times more mass in the epoch prior to the
Late Heavy Bombardment, real time would be $0.25 \times 10^{6}t'$, so that
800Myr corresponds to $t'=3200$, and again we would not expect collisional
evolution to be strong.
However, it must also be noted that we have assumed here that planetesimals remain
in their $(q,Q)$ cell and are only removed by collisions.
In the scattered disk dynamical processes, such as scattering by Neptune, can occur
on timescales that are shorter than collisional mass loss, and so must be taken
into account (e.g., Volk \& Malhotra 2008), although the 2003 EL$_{61}$ collisional
family in the Kuiper belt may provide evidence of the role of collisions
in the evolution of the scattered disk (Levison et al. 2008).

Another application would be to a putative scattered disk around HD69830.
For a pericentre at 1AU, just outside the outermost (known) planet (Lovis et al. 2006),
we see that for an initial mass of $1M_\oplus$ real time is the same as $t'$
(for $K=2.1 \times 10^{-14}$ and $I_{\rm{max}}=0.05$), and so after $\sim 2$Gyr of evolution the 
maximum mass that can remain is independent of both initial mass and $\alpha$, although in 
accordance with the earlier discussion it does depend on the maximum eccentricity in the 
distribution.
Thus we find that $10^{-3}M_\oplus$ can remain at 2Gyr if apocentres extend out to
50AU, with more mass remaining should higher apocentres be present (or lower values
of $K$ be applicable).
It is worth noting from Fig.~\ref{fig:sd}b, however, that although we have increased
the mass that passes through 1AU, this does not necessarily increase the mass surface
density at 1AU.

%%%%%%%%%%%%%%%%%%%%%%%%%%%%%%%%%%%%%%%%%%%%%%%%%%%%%%
%%%%%%%%%%%%%%%%%%%%%%%%%%%%%%%%%%%%%%%%%%%%%%%%%%%%%%
\section{Emission properties of eccentric planetesimal swarms}
\label{s:3}
It was shown in \S \ref{s:2} and \S \ref{s:4} how increasing planetesimal
eccentricity leads to longer collisional lifetimes, and higher disk 
masses at late times in spite of collisional processing.
Here we consider how the emission spectrum of a planetesimal swarm changes
as its eccentricity is increased and how such swarms might appear in
observations of dust around nearby stars.

%%%%%%%%%%%%%%%%%%%%%%%%%%%%%%%%%%%%%%%%%%%%%%%%%%%%%%
\subsection{Emission spectrum}
The emission spectrum from planetesimal belts comprised of dust with
absorption (and emission) efficiencies $Q_{\rm{abs}}(\lambda,D)$ can
be calculated using:
\begin{eqnarray}
  F_\nu & = & 2.35 \times 10^{-11} d^{-2} \sigma_{\rm{tot}} 
              \sum_r \bar{\sigma}(r) \int_{D_{\rm{min}}}^{D_{\rm{max}}}
              Q_{\rm{abs}}(\lambda,D) \times \nonumber \\
        &   &    
                 B_\nu(\lambda,T(D,r))
                 \bar{\sigma}(D)dD, \\
  T(D,r)  & = &  [ \langle Q_{\rm{abs}}(D,\lambda) \rangle_{T_\star} /
                   \langle Q_{\rm{abs}}(D,\lambda) \rangle_{T(D,r)} ]^{1/4}
                   T_{\rm{bb}}(r), \\
  T_{\rm{bb}}(r) & = & 278.3 L_\star^{0.25} r^{-0.5},
\end{eqnarray}
where $F_\nu$ is in Jy, $d$ is distance in pc, $\sigma_{\rm{tot}}$ is in AU$^2$ and scales with 
total mass according to equation~(\ref{eq:stot}) for the assumptions of this paper about the size 
distribution, $\bar{\sigma}(D)dD$ is the fraction of the total cross-sectional area in the
size range $D$ to $D+dD$ and is $\propto D^{-1.5}dD$ for the distribution assumed 
here, $D_{\rm{min}}$ is the minimum size in the distribution
that is commonly assumed to be the size at which dust is blown out of the system
by radiation pressure, and $D_{\rm{max}}$ is the size of the largest object, although
for computational purposes this can be set at $\sim 1$m as larger objects contribute
little to the emission spectrum (e.g., Fig. 5 of Wyatt \& Dent 2002),
$B_\nu$ is in Jy sr$^{-1}$, $\langle Q_{\rm{abs}} \rangle_T$ means $Q_{\rm{abs}}$ averaged over a 
black body spectrum of temperature $T$, and $L_\star$ is in $L_\odot$.

Later in the paper we will use emission efficiencies calculated using Mie theory
(or another suitable approximation) along with the optical constants of different
materials combined using a mixing theory in the manner described elsewhere
(Li \& Greenberg 1998; Wyatt \& Dent 2002).
The grain model we will use assumes a core-mantle composed of silicates and organic refractory
material, in the ratio $q_{\rm{Si}}$, which is incorporated into the grain with a
porosity $p$;
some fraction $q_{\rm{H_2O}}$ of the vacuum is filled with water ice.
However, for heuristic purposes this section starts by assuming that
dust created in the planetesimal belt both absorbs and emits light
like a black body so that $Q_{\rm{abs}}=1$.

\begin{figure}
  \begin{center}
    \vspace{-0.2in}
    \begin{tabular}{cc}
      \textbf{(a)} & \hspace{-0.5in} \psfig{figure=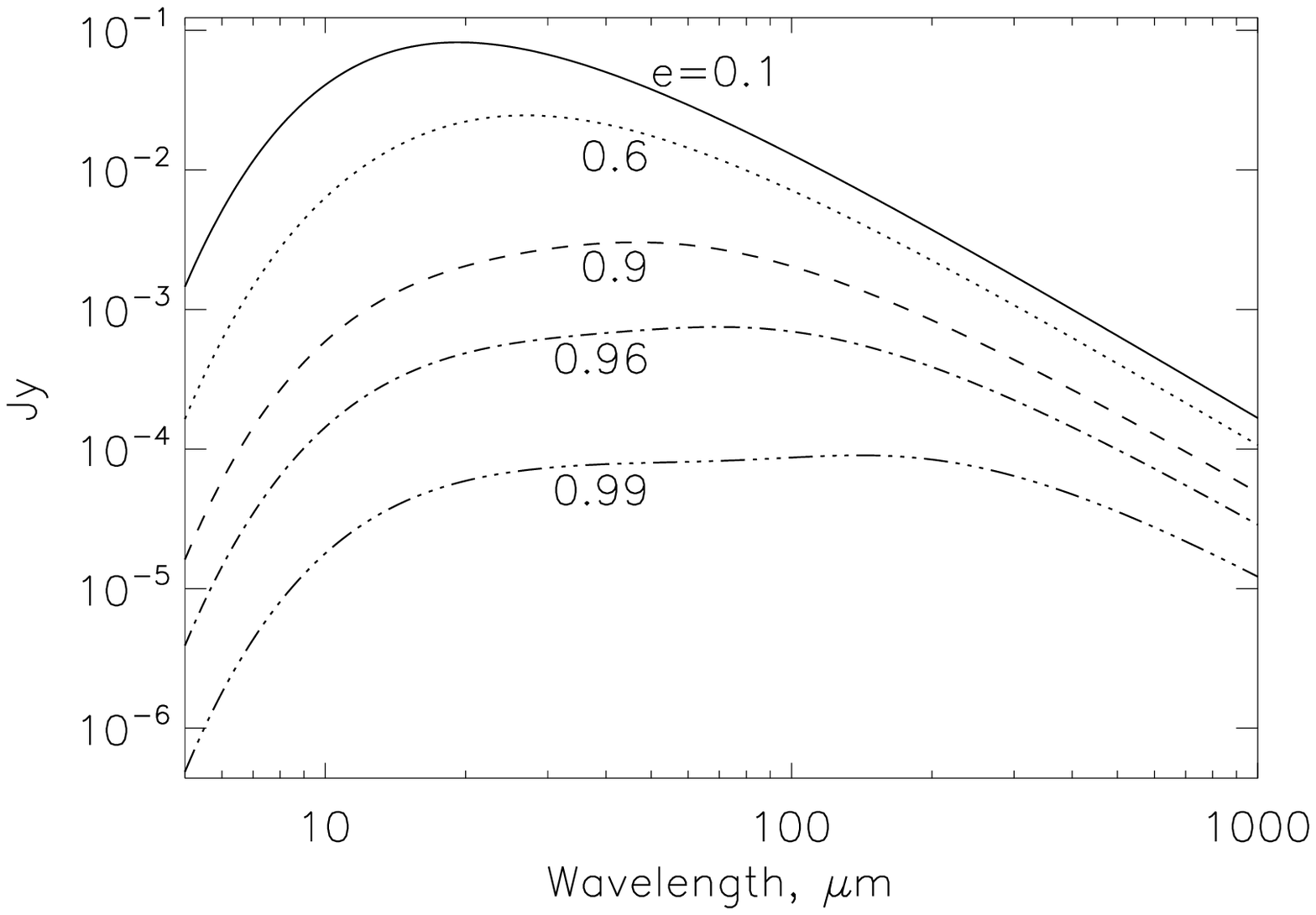,height=2.4in} \\
      \textbf{(b)} & \hspace{-0.5in} \psfig{figure=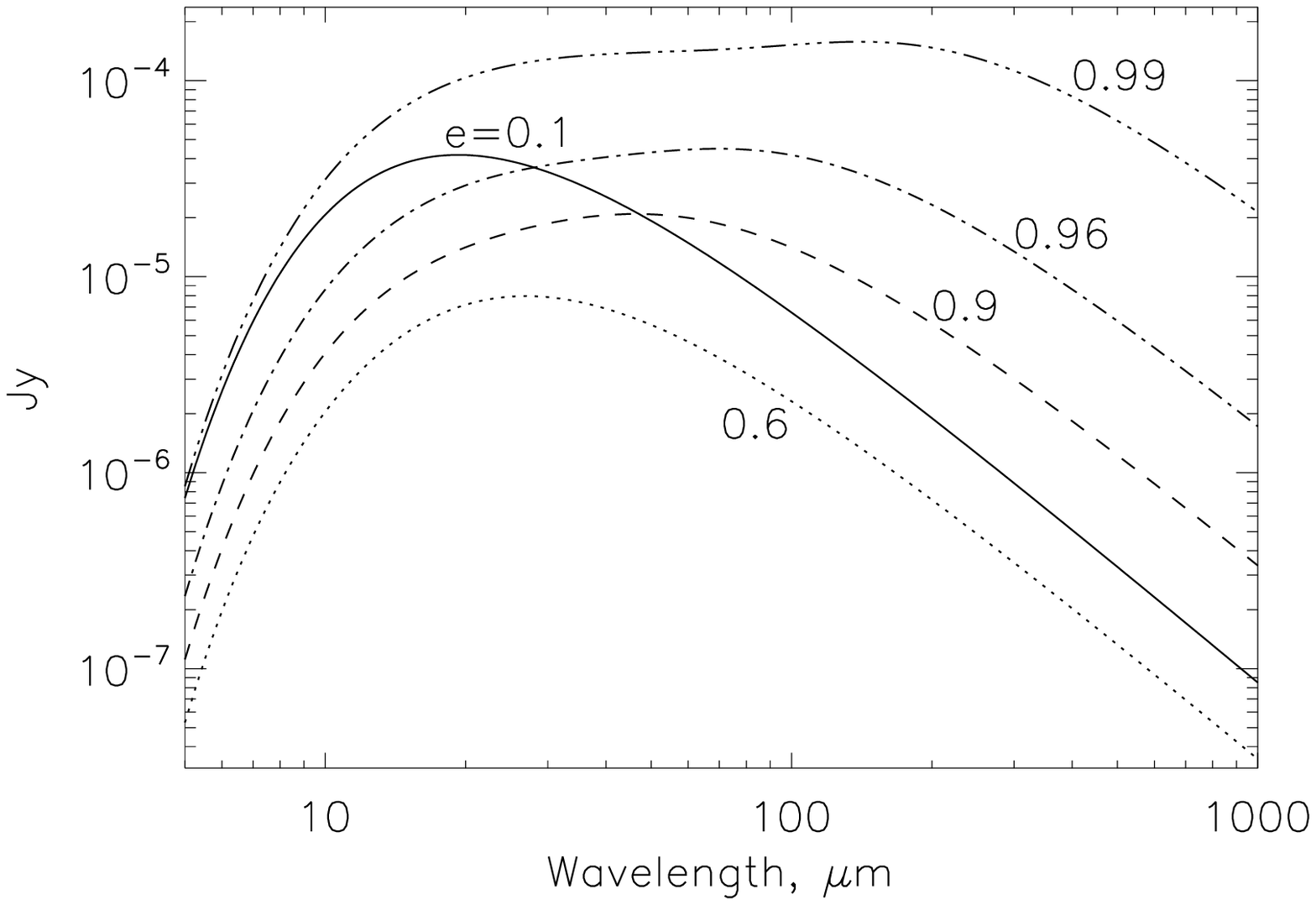,height=2.4in} \\
      \textbf{(c)} & \hspace{-0.5in} \psfig{figure=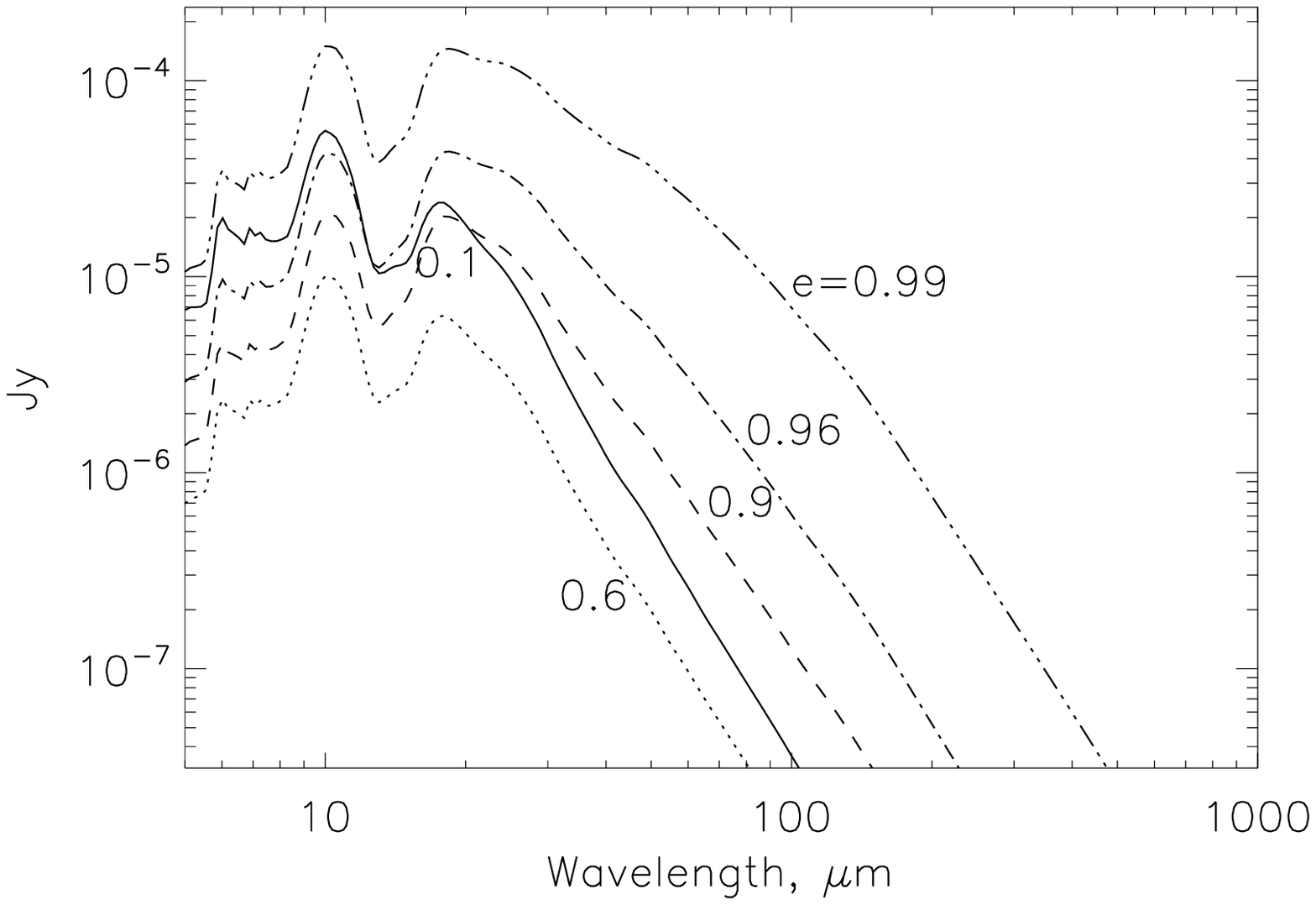,height=2.4in}
    \end{tabular}
    \caption{Emission spectra of disks comprised of material all with the same
    pericentres, but with eccentricities of 0.1, 0.6, 0.9, 0.96, 0.99 shown with different 
    linestyles. 
    \textbf{(a)} Spectra for $10^{-3}$AU$^2$ of material with black-body emission 
    properties and with pericentres at 1AU around a $1L_\odot$ star at 10pc.
    \textbf{(b)} As for \textbf{(a)}, but the total cross-sectional area has been scaled to
    the maximum mass that can remain after 1Gyr of processing
    (assuming $K=2.1 \times 10^{-14}$ and $I_{\rm{max}}=0.05$),
    also taking into account the change in blow-out radius as eccentricity increases.
    \textbf{(c)} As for \textbf{(b)}, but assuming the cross-sectional area is in 1 $\mu$m
    grains of a mixture of silicate-organic refractory material.}
   \label{fig:bbsed}
  \end{center}
\end{figure}

For the disk comprised of planetesimals with common pericentre and apocentre distances 
discussed in \S \ref{s:2}, Figure~\ref{fig:bbsed}a shows the emission spectrum for a range 
of eccentricities for a star with $L_\star=1L_\odot$, $d=10$pc,
and for a planetesimal belt with $q=1$AU and $\sigma_{\rm{tot}}=10^{-3}$AU$^{2}$.
Although the absolute level of emission and range of wavelengths are dependent on these
stellar and planetesimal belt properties, the shape of the emission spectrum would not be
(e.g., wavelengths would scale $\propto L_\star^{-0.25} q^{0.5}$).
This shows that increasing eccentricity leads both to emission over a wider range of wavelengths,
due to the larger range of radii and so temperatures in the disk, and to a decrease in the level
of emission for a given amount of cross-sectional area.

To work out the emission spectra for these disks at late times, equation~\ref{eq:mlate2}
was used to get the mass, and equation~\ref{eq:stot} to get the cross-sectional area
(see Fig.~\ref{fig:bbsed}b).
The minimum size in the distribution was assumed to be the size at which dust is blown-out
by radiation pressure.
Because the majority of collisions occur at pericentre (see Figure~\ref{fig:sevsr}c),
the blow-out size is that for which radiation pressure $\beta=0.5(1-e)$, so that for
black body grains
\begin{equation}
  D_{\rm{bl}} = 0.8(2700/\rho)(L_\star/M_\star)/(1-e), \label{eq:dbl}
\end{equation}
in $\mu$m.
The increase in blow-out size for high eccentricities is significant since it
truncates the collisional cascade thus removing much of its cross-sectional
area.
Figure~\ref{fig:bbsed}b shows that the effect of increasing 
eccentricity of a planetesimal belt above 0.1 while keeping its pericentre distance constant
is first to reduce the amount of emission present at late times at all wavelengths
(compare the $e=0.1$ and $e=0.6$ lines).
However, for high enough eccentricities the emission is increased above the $e=0.1$
values, with the transition occurring at eccentricities of $\sim 0.99$ for short wavelength
emission and $\sim 0.7$ for long wavelength emission.

Although the quantitative conclusions about the absolute level of emission would be
different if realistic particles had been assumed, these qualitative conclusions
about how the emission spectrum changes as eccentricity is changed would not, 
since a change in particle properties would affect the emission from all disks in a
similar manner.
To illustrate this, Fig.~\ref{fig:bbsed}c shows the emission spectra assuming that the
total cross-sectional area is all in 1 $\mu$m particles of silicate-organic refractory
material (noting that this is not meant to be a physical model).
The emission efficiencies of such particles drops rapidly at longer wavelengths (apart
from close to the 10 and 18 $\mu$m silicate features) dramatically reducing the far-IR
emission, even though the qualitative comparison of the behaviour as eccentricity is
changed is unaffected (i.e., the eccentricity required for an increase in short or
long-wavelength emission is the same).

%%%%%%%%%%%%%%%%%%%%%%%%%%%%%%%%%%%%%%%%%%%%%%%%%%%%%%
\subsection{Consequences of increased remaining mass}
\label{ss:cons}
While increasing the eccentricity does not have the effect of increasing the warm disk
emission, it does increase the remaining mass, and that has two important implications.

%%%%%%%%%%%%%%%%%%%%%%%%%%%%%%%%%%%%%%%%%%%%%%%%%%%%%%
\subsubsection{Blow-out population}
\label{sss:blowout}
The first is that the collisional cascade is losing mass through radiation
pressure blow-out at a rate 
$M_{\rm{tot}}/t_{\rm{cc}}(D_{\rm{max}})$, which is
$M_{\rm{late}}/t_{\rm{age}}$ once the largest objects reach collisional equilibrium.
This means that a higher remaining mass means a higher mass loss rate.
Typically the short lifetime of blow-out grains means that they contribute little to the
total cross-sectional area present in the disk.
However, for an eccentric ring in which the majority of the mass loss occurs at
pericentre, the surface density of the blow-out grains can exceed that of the collisional 
cascade.

The following arguments give an estimate of the surface brightness of these grains.
To simplify this calculation it is assumed that all collisions occur at pericentre where
dust of size $D_{\rm{loss}}$ (in $\mu$m) is released at the orbital velocity of
$v_{\rm{loss}} = 2\pi \sqrt{M_\star(1+e)/q}$ AU yr$^{-1}$.
Further assuming that this dust has $\beta=1$ means that the dust undergoes no acceleration
so that
\begin{equation}
  \dot{r}_{\rm{loss}} = v_{\rm{loss}} \sqrt{1-(q/r)^2}.
  \label{eq:rdotloss}
\end{equation}
The mass loss rate, when converted to a rate of loss of cross-sectional area as 
dust of size $D_{\rm{loss}}$ and divided by the cross-sectional area present in the
collisional cascade gives
\begin{equation}
  \dot{\sigma}_{\rm{loss}}/\sigma_{\rm{totcc}} = 3.16 \times 10^4
    D_{\rm{bl}}^{0.5} D_{\rm{max}}^{0.5} D_{\rm{loss}}^{-1} t_{\rm{age}}^{-1},
\end{equation}
which results in a distribution of cross-sectional area for $r>q$ of
\begin{eqnarray}
  d\sigma_{\rm{loss}}/\sigma_{\rm{totcc}}/d\bar{r} & = &
     4500 L_\star^{0.5}M_\star^{-1}D_{\rm{max}}^{0.5}
     D_{\rm{loss}}^{-1}t_{\rm{age}}^{-1} q^{1.5}
     \times \nonumber \\ & &
     (1+e)^{-0.5}(1-e)^{-1.5} [1-(q/r)^2]^{-0.5}.
\end{eqnarray}

This can be compared with the distribution of cross-sectional area in the collisional
cascade (Fig.~\ref{fig:sevsr}a), which the analytical results (eq.~\ref{eq:sigbaranal})
show has a minimum at $\bar{r}=1-e^2$ of $\bar{\sigma}/d\bar{r}=\pi^{-1}\sqrt{e^{-2}-1}$.
Thus the blow-out population becomes more important relative to the collisional cascade
as eccentricity increases, and is also more dominant in populations with larger pericentre 
distances.
Specifically, the cross-sectional area in the blow-out population can be higher than
that of the collisional cascade at some radius when
\begin{equation}
  e > 1 - 160L_\star^{0.25}M_\star^{-0.5}q^{0.75}D_{\rm{max}}^{0.25}
    D_{\rm{loss}}^{-0.5}t_{\rm{age}}^{-0.5}
\end{equation}
for these assumptions.
For the $M_\star=L_\star=q=1$ system at 1Gyr considered earlier,
further assuming that $D_{\rm{max}}=2000$km and $D_{\rm{loss}}=1\mu$m,
this means that the blow-out population becomes important when $e>0.97$,
noting that $D_{\rm{loss}} \ll D_{\rm{bl}} = 30\mu$m in this example.

%%%%%%%%%%%%%%%%%%%%%%%%%%%%%%%%%%%%%%%%%%%%%%%%%%%%%%
\subsubsection{Cometary sublimation}
\label{sss:cometsub}
The conclusions of \S \ref{sss:blowout} are independent of the mechanism
producing the blow-out grains, which need not be through collisions.
Rather the large remaining mass might provide a reservoir that, through
collisions, replenishes smaller planetesimals that release dust through
comet-like sublimation.
In such a case the above calculation may overestimate the amount of
cross-sectional area in the collisional cascade meaning that the
cascade could be extremely faint, because sublimation timescales
(which are $\propto D$ if mass loss rate is proportional to
surface area) can be shorter than collision timescales (which from
eq.~\ref{eq:rcc} are $\propto D^{0.5}$) for the small
particles in the cascade, thus reducing their number so that 
$D_{\rm{min}}$ should be significantly higher than $D_{\rm{bl}}$
in equation~(\ref{eq:stot}).
However, this would only be the case if sublimation processes just
produce non-sublimating (e.g., silicate) grains that are below the
blow-out limit.
Large non-sublimating grains produced above the blow-out limit
would increase the cross-sectional area in the collisional cascade.
Since Solar System comets are seen to release significant quantities of
non-icy mm-cm-sized grains (e.g., Reach et al. 2009), we consider that further
work is required before the impact of sublimation on the steady-state size
distribution of a collisional cascade is fully understood, but note that
sublimation remains a viable mechanism for feeding the blow-out population
discussed in \S \ref{sss:blowout}.

%%%%%%%%%%%%%%%%%%%%%%%%%%%%%%%%%%%%%%%%%%%%%%%%%%%%%%
\subsubsection{Frequency of massive collisions}
\label{sss:massivecoll}
To estimate the frequency of massive collisions for a
system evolving in collisional equilibrium, for which the collisional
lifetime of the largest objects is the age of the system, we rewrite
equation~24 of Wyatt et al.~(2007a) to find that the collision
rate for objects larger than size $D_{\rm{pb}}$ is given by
\begin{equation}
  R_{\rm{cc}}(D>D_{\rm{pb}}) = (M_{\rm{late}}/6M_{\rm{pb}})t_{\rm{age}}^{-1},
  \label{eq:rccdpb}
\end{equation}
where a distribution with $q=11/6$ was assumed.
This means that, for a given parent body size, a higher eccentricity ring 
results in a higher remaining mass and so more frequent collisions.

However, a higher eccentricity also means that larger parent bodies are
required to reproduce the same fractional luminosity.
To estimate the minimum mass of a parent body, $D_{\rm{pb}}$, that would
be required to be destroyed to reproduce an observed fractional
luminosity of $f_{\rm{obs}}$, it is assumed (optimistically) that a collision
turns all of an object's mass into particles of size $D_{\rm{bl}}$ which are
then redistributed around a ring with the same eccentricity as the
parent object (i.e., ignoring the increase in eccentricity due to
radiation pressure).
The fractional luminosity from an eccentric ring can be calculated by
integrating equation (\ref{eq:sigbaranal}) to be
\begin{equation}
  f_{\rm{obs}} = \sigma_{\rm{tot}} / (4\pi q^2 (1-e)^{-2} \sqrt{1-e^2}).
\end{equation}
Thus to reproduce $f_{\rm{obs}}$ requires a parent body mass
\begin{equation}
  M_{\rm{pb}} = 3.14 \times 10^{-8} \rho D_{\rm{bl}} f_{\rm{obs}} q^2 (1-e)^{-2} (1-e^2)^{0.5}
\end{equation}
in $M_\oplus$, noting that $D_{\rm{bl}}$ is also a function of eccentricity.
This readily gives the maximum rate of collisions that
could reproduce a given fractional luminosity in a single event;
e.g., if a mass of $1M_\oplus$ remains at 1Gyr in a belt with a pericentre
of 1AU and eccentricity of 0.99, then events that might potentially produce 
$f_{\rm{obs}}=10^{-3}$ in 1$\mu$m dust occur 1.4 times per Myr.

On the other hand a higher eccentricity means that the dust is not depleted in
collisions so rapidly.
The collision rate of same sized particles can be worked out using
equation~(\ref{eq:rcc}) with $f_{\rm{cc}}(D)=4$ (Wyatt et al. 1999).
Using the analytical approximation of \S \ref{sss:analytical} we find that
\begin{equation}
  R_{cc}(D_{\rm{bl}}) = 1.3 \sigma_{\rm{tot}} M_\star^{0.5} q^{-3.5}
     (1-e)^{3.5} (1-e^2)^{-0.5} I_{\rm{max}}^{-1},
\end{equation}
which is true regardless of particle size (as long as such particles contain the
majority of $\sigma_{\rm{tot}}$), and gives a collision rate
to reproduce $f_{\rm{obs}}$ of
$16f_{\rm{obs}}M_{\rm{obs}}^{0.5}I_{\rm{max}}^{-1}a^{-1.5}$.

Combining these results in the same manner as Wyatt et al. (2007a), to estimate
the fraction of time that collisions are expected to result in dust above
a given level of $f_{\rm{obs}}$, we find that
\begin{eqnarray}
  P(f>f_{\rm{obs}}) & = & 0.33 \times 10^{6} M_{\rm{late}} t_{\rm{age}}^{-1} 
     f_{\rm{obs}}^{-2} M_\star^{-0.5} I_{\rm{max}}
  \times \nonumber \\ & & 
  \rho^{-1} D_{\rm{bl}}^{-1} q^{-0.5} (1+e)^{-0.5}.
\end{eqnarray}
For example, if a mass of $1M_\oplus$ remains at 1Gyr in a belt with a pericentre
of 1AU and eccentricity of 0.99, then events that might potentially produce 
$f_{\rm{obs}}=10^{-3}$ in 1$\mu$m dust of density 2700 kg m$^{-3}$ with inclinations
up to 0.05, would occur 0.4\% of the time.
This fraction is insensitive to eccentricity except that a high value is required to 
permit $1M_\oplus$ of material to remain so late despite collisional processing, and 
would also affect the validity of assuming the dust is 1$\mu$m in size.

%%%%%%%%%%%%%%%%%%%%%%%%%%%%%%%%%%%%%%%%%%%%%%%%%%%%%%
\subsection{Application to $\eta$ Corvi}
\label{ss:etacorvi}
The 1.3 Gyr F2V star $\eta$ Corvi at 18.2pc exhibits excess emission from 
circumstellar dust at wavelengths from a few $\mu$m up to sub-mm.
The sub-mm emission has been resolved in imaging at both 450 and 850$\mu$m with SCUBA,
and modelling shows this dataset can be explained by emission from a ring of $150 \pm 20$ AU 
radius inclined at $45 \pm 25^\circ$ to our line-of-sight (Wyatt et al. 2005).
The region is seen to be centrally cleared, but the excess mid-IR emission cannot 
originate in the 150AU ring and has been resolved to lie somewhere in the 0.16-3.0AU
region (Smith et al. 2008; Smith, Wyatt \& Haniff 2009b) with a temperature
inferred from the emission spectrum that places it at $\sim 1.7$AU.
It remains debatable whether there is any emission in the 3-100AU region (Chen et al. 
2006; Smith et al. 2008).
Regardless the warm emission at $<3$AU is expected to be transient if confined to
a ring at that location (Wyatt et al. 2007a), as
discussed in the introduction.

\begin{figure}
  \begin{center}
    \vspace{-0.2in}
    \begin{tabular}{c}
%      \textbf{(a)} & \hspace{-0.5in}
        \psfig{figure=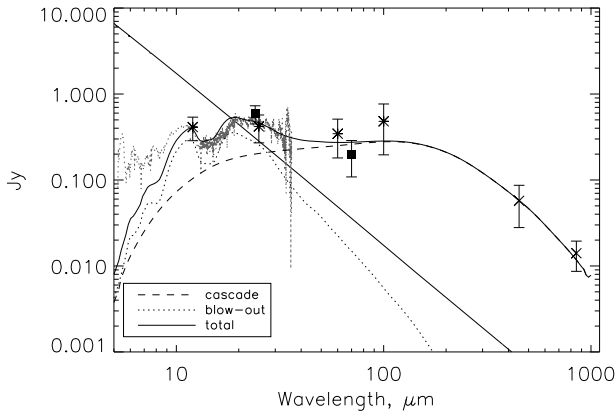,height=2.4in}
    \end{tabular}
    \caption{Emission spectrum of $\eta$ Corvi modelled using a single planetesimal
    population with pericentres at 0.75AU and apocentres at 150AU (see text for details).
    The dashed, dotted and solid lines correspond to the contribution from the
    collisional cascade, from blow-out grains, and the total emission spectrum,
    respectively.
    The diagonal solid line is the stellar spectrum, the grey dotted line is the IRS
    spectrum after the photospheric contribution has been subtracted, asterisks
    and squares are the IRAS and Spitzer photometric fluxes, respectively, after 
    photosphere subtraction (and colour correction where necessary).}
   \label{fig:etacorvi}
  \end{center}
\end{figure}

Here we consider the possibility that both the imaging and spectroscopic constraints can
be explained by a model with a single planetesimal population.
Figure~\ref{fig:etacorvi} shows the emission spectrum of a model with planetesimals of 
density 1480 kg m$^{-3}$ with pericentres at 0.75AU, apocentres at 150AU (i.e., 
eccentricity 0.99), and inclinations up to 0.05, after 1.3Gyr of evolution.
The planetesimal collisional parameters are assumed to be $D_{\rm{max}}=2000$km with 
strength $Q_{\rm{D}}^\star=3 \times 10^{5}$ J kg$^{-1}$, leading to the conclusion that 
$5M_\oplus$ of material remains at the current epoch.
Dust is created in the collisional cascade down to a size of $D_{\rm{bl}}=480$ $\mu$m
before removal by radiation pressure, and the emission properties of the grains were 
calculated assuming a core-mantle composed of 30\% amorphous silicate and 70\% organic 
refractory material with porosity of 0.4 and water ice filling 20\% of the gaps;
the emission spectrum of the collisional cascade is shown with a dashed line,
and is largely insensitive to the assumed composition as bound grains act
like black bodies.
The population of grains that are being removed by radiation pressure are assumed
to be all of 5$\mu$m in diameter, and their emission properties were calculated 
assuming the same grain properties as the collisional cascade;
their emission spectrum is shown with a dotted line.

The model provides an excellent fit to the observed shape of the
emission spectrum, noting that it is not our intention to provide an exact fit
to the emission features for which a more detailed compositional model would be 
required; i.e., the composition should not be regarded to be constrained
in this model, but in providing a fit to the spectrum it does
allow the physical parameters of the model to be self-consistently derived.
In the model, the mid-IR emission is dominated by grains being removed by
radiation pressure, which contribute 91\% and 84\% of the 11.6$\mu$m and
18.7$\mu$m emission, respectively.
Such emission is centrally concentrated (e.g., 68\% and 51\% of the emission
in these wavebands comes from inside 3.5AU).
To assess whether the model is consistent with available mid-IR imaging constraints
we first show that the model fluxes in an aperture of radius 0.5\arcsec
(285mJy and 263mJy at these wavebands) are consistent with those
observed ($330 \pm 184$mJy and $309 \pm 79$mJy, Smith et al. 2008).
In the model, the aperture used to estimate the background level, 0.5-1.0\arcsec,
does include disk emission (28mJy and 60mJy in these wavebands), but
at a level consistent with its non-detection in Smith et al. (2008) for which
$>150$ and $>140$mJy would have been necessary for a $>3\sigma$ detection.
To consider whether there is an aperture which could confidently detect the
mid-IR emission from such a model we plotted the model fluxes on Fig.~7 of Smith
et al. (2008), noting that the central concentration of the mid-IR emission means
that we only need to consider the emission close to the pericentre.
Thus we approximated the mid-IR emission from the model as originating in
two consecutive annuli each with a uniform surface brightness and 
spanning radii $3 \pm 3$AU and $9 \pm 3$AU;
from the model we find that the inner annulus contains 262mJy and 239mJy, and
the outer annulus contains 27mJy and 53mJy at these wavebands.
Both annuli are below the threshold at which extended emission should have been 
detected, consistent with the observations, but the outer annulus is at a level where 
extension would be detected in observations twice as deep at both wavebands.
The lack of material $<0.75$AU is consistent with visibilities
measured by MIDI that suggest the emission is completely resolved,
although the model would predict small changes in
visibility with baseline for a 40mas ring that might be detectable in more
sensitive observations (see Fig.~9 of Smith et al. 2009b).
The low emission efficiencies of the blow-out population means that it
contributes little to the longer wavelength fluxes, and the sub-mm emission is 
dominated by the collisional cascade.
The majority of the sub-mm emission comes from material near apocentre
(92\% of the 450 $\mu$m emission comes from $100-150$AU) and so we
expect the model to fit the sub-mm imaging constraints (Wyatt et al. 2005).

Although we have devised a model that explains the observations, this
does not necessarily make it plausible.
Nevertheless, most of the model parameters are reasonable.
The planetesimal strength required is above that assumed for the
population models of Wyatt et al. (2007b), but is within the range of
that expected for 2000km planetesimals (Benz \& Asphaug 1999).
The assumption that blow-out particles are 100 times smaller
than the blow-out limit could indicate that larger unbound particles
readily disintegrate into smaller fragments (noting that they are produced at
collisional velocities up to 80 km s$^{-1}$), but may perhaps be circumvented
with a different choice of composition.
It is also worth noting that this model predicts that we have a 1:2300
chance of witnessing a collision capable of reproducing the
observed fractional luminosity of $0.5 \times 10^{-3}$;
i.e., it is possible that the mid-infrared emission is enhanced due to
stochastic collisions.

The origin of the planetesimals is, however, a concern.
High eccentricity orbits normally imply scattering by a planet, with the
apocentre or pericentre coinciding with the orbit of the planet.  
It is unclear why both pericentre and apocentre distances would be fixed here.
One possibility is that a primordial population that included a wider range of 
eccentricities and semimajor axes has been depleted by
collisional processing leaving just the high eccentricity remnant
(see \S \ref{ss:sd}).
The primordial population could then be an extended scattered disk
created by inward planet migration (Payne et al. 2009), with a planet
lying just interior to the pericentre, and the apocentre corresponding
to the highest eccentricity attained in the population.
In this case the current mass is a small fraction of the primordial mass;
e.g., for $n(Q) \propto Q^{-\alpha}$ where $\alpha=1$ or 2 we find the primordial
mass to be 7 or 200 times (respectively) the $5M_\oplus$ inferred at present
(assuming the observed apocentres are in the range 90-210AU), and so an
initially flat apocentre distribution is required for a realistic starting
population.
Alternatively this population could have been scattered in by a more distant
planet which was orbiting close to the apocentre, although in this scenario
the lack of material with pericentres in the 3-100AU range is a problem, as
collisional processing preferentially removes material with the lowest
pericentres in a distribution with a common apocentre.
\footnote{Collision rates for material with common apocentres, $Q_1$, would
be given by equation~(\ref{eq:rccs2}) replacing $q_1^{-13/3}$ with
$Q_1^{-13/3}[(1+e_1)/(1-e_1)]^{13/3}$.}
Certainly the dynamics of creating populations of extremely high eccentricity
must be explored before we can be confident that this is a viable model.
It is also important that the dynamics that creates the high planetesimal
eccentricities, such as scattering by a planet, does not 
significantly deplete the planetesimal population on longer timescales.
This could be achieved if the planetesimals become detached from the 
planet, perhaps because that planet migrates or is scattered to put it
out of reach of the planetesimals.

Regardless of the origin of the planetesimals, it seems that the
0.75-150AU region would have to be clear of planets for this model to be viable.
Current limits from radial velocity studies only rule out companions down to
2.1 Jupiter mass out to 0.48AU (Lagrange et al. 2009).
As noted above, another testable prediction of the model is that there should
be thermal emission from the 3-100AU region which may be detectable as
extended emission.

%%%%%%%%%%%%%%%%%%%%%%%%%%%%%%%%%%%%%%%%%%%%%%%%%%%%%%
\subsection{Application to HD69830}
\label{ss:hd69830}
The 2Gyr K0V star HD69830 at 12.6pc exhibits excess emission from circumstellar
dust at mid-infrared wavelengths, including significant emission features
(Beichman et al. 2005).
No excess emission is seen at 70$\mu$m (Trilling et al. 2008) or at
850$\mu$m (Sheret, Dent \& Wyatt 2004; Matthews, Kalas \& Wyatt 2007).
Three Neptune-mass planets are known within 0.7AU (Lovis et al. 2006)
that are thought to have migrated in from 3-8AU (Alibert 
et al. 2006).
Mid-infrared studies constrain the dust to 0.05-2.4AU (Smith et al. 2009b),
consistent with the temperature inferred from the emission spectrum that places
the dust at 1AU (Lisse et al. 2007);
at such a location the dust is inferred to be transient (Wyatt et al. 2007a).

\begin{figure}
  \begin{center}
    \vspace{-0.2in}
    \begin{tabular}{c}
%      \textbf{(a)} & \hspace{-0.5in}
        \psfig{figure=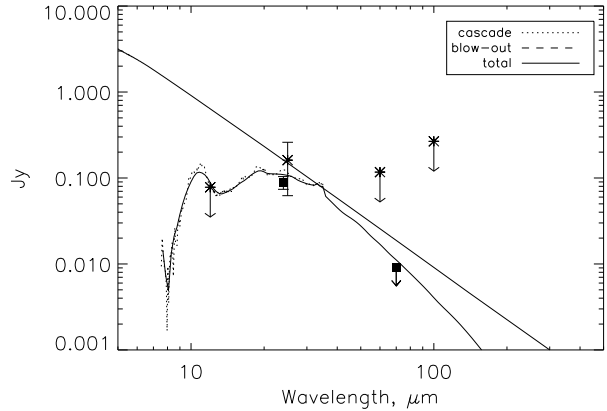,height=2.4in}
    \end{tabular}
    \caption{Emission spectrum of HD69830 modelled using a single planetesimal
    population with pericentres at 1.2AU and eccentricity 0.99 (see text for 
    details).
    The solid line shows the contribution from blow-out grains;
    for this to be a viable model the emission from the collisional cascade
    component has been suppressed, e.g., due to the sublimation of its smallest grains. 
    The diagonal solid line is the stellar spectrum, the grey dotted line is the IRS
    spectrum after the photospheric contribution has been subtracted,
    asterisks and squares are the IRAS and Spitzer photometric fluxes, respectively, 
    after photosphere subtraction (and colour correction where necessary).}
   \label{fig:hd69830}
  \end{center}
\end{figure}

The lack of cold emission rules out several models for the
origin of the dust, including the eccentric planetesimal belt model
as applied to $\eta$ Corvi above.
Although an eccentric ring alleviates concerns about the longevity
of the disk feeding the dust at 1AU, the hot dust is always accompanied by
cold emission from collisional cascade material at apocentre (Fig.~\ref{fig:bbsed}b).
One way around this is to postulate that there is little cross-sectional
area in the collisional cascade, e.g. because small grains have been
removed by sublimation (\S \ref{sss:cometsub}).
Figure~\ref{fig:hd69830} shows a model which fits the emission
spectrum by only including the blow-out component comprised of solid grains 1$\mu$m
in diameter composed of 1/3 amorphous silicate and 2/3 organic refractory material;
note again that the model is used only to fit the overall shape of the 
spectrum, and does not claim the level of detail required to constrain the 
composition.
Dust in this model extends out from the 1.2AU pericentre of a ring with
eccentricity 0.99;
the spectrum is insensitive to eccentricity, however this does have a moderate
impact on the inferred mass loss rate through equation~(\ref{eq:rdotloss}).
As well as nearly fitting the 70$\mu$m upper limit, this model also meets the mid-infrared 
imaging constraints, since it predicts the 18.7$\mu$m flux within a 0.5\arcsec 
aperture to be 412mJy (compared with $377 \pm 46$mJy found by Smith et al. 2009b),
with just 30mJy coming from the 2-4\arcsec region.
As for \S \ref{ss:etacorvi} the model was analysed assuming the emission to 
originate in two annuli at $2 \pm 2$AU and $6 \pm 2$AU with disk fluxes of
121 and 29mJy, respectively.
Comparison with Fig.~7 of Smith et al. (2009b) shows that both annuli are below the 
threshold at which extended emission should have been detected in these 
observations, but that extension would have been detected from the outer annulus in 
observations twice as deep, a conclusion which holds regardless of the
origin of the dust as long as it is assumed to be in the process of radiation
pressure blow-out.
Without material $<1.2$AU the model emission would be expected to be
completely resolved on MIDI baselines in line with observations
(Smith et al. 2009b).

This model gives a mass loss rate of $0.08M_\oplus$ Myr$^{-1}$, which
is higher than the $5-60\times 10^{-6}M_\oplus$/Myr quoted in
Beichman et al. (2005) and Wyatt et al. (2007a), since those papers
assumed mass loss due to collisions rather than radiation pressure.
Assuming this mass loss comes from a parent population in collisional
equilibrium, this implies a parent population of $160M_\oplus$, 
similar to the $90-900M_\oplus$ calculated in Beichman et al.
(2005) based on an extrapolation from Hale-Bopp mass loss rates. 
It seems that such a high remaining mass is prohibitive after 2Gyr
of processing, since this can only be achieved for $e>0.998$ for the
fiducial value of $K=2 \times 10^{-14}$ or $e>0.993$ for the slightly higher
K-value used for $\eta$ Corvi modelling (\S \ref{ss:etacorvi}).
Even if it was possible to implant such a large mass at high eccentricities, there would
still be the concern raised in \S \ref{sss:cometsub} about whether comet sublimation 
would also produce bound non-sublimating grains that would increase the cold emission 
from the collisional cascade.
Thus a steady-state explanation is not favoured.

An alternative model invokes a recent collision as the origin of the dust.
In an eccentric ring such events can occur frequently enough for this model
to be viable (\S \ref{sss:massivecoll}).
A planetesimal belt mass of $1M_\oplus$ results in $P(f>0.2 \times 10^{-3}) = 0.06$ 
assuming that collisions result in $1\mu$m grains.
Since the observed grains also seem to be small enough to be blown out by radiation
pressure, the probability of witnessing such an event is in fact lower than that
derived in \S \ref{sss:massivecoll} and quoted above:
a short grain lifetime either implies the collision occurred very recently
(i.e., within the blow-out timescale of several years), or that the dust
is the product of ongoing secondary collisions or sublimation amongst the
debris of an older more massive (and hence rarer) collision.
The persistence of 24 $\mu$m excess over 24 year timescales between the epochs of IRAS
and Spitzer (Lisse et al. 2007) rules out
a very recent origin.
This leaves ongoing secondary collisions or sublimation in a debris belt which
would have to contain $>2 \times 10^{-6}M_\oplus$ to sustain a mass loss of
$0.08M_\oplus$ Myr$^{-1}$ over 24 years.
Equation~(\ref{eq:rccdpb}) shows that such debris belts might only be recreated in
collisions every 24,000 years for a planetesimal belt containing $1M_\oplus$ at 2Gyr.
In other words, we have a 1:1000 chance of witnessing the aftermath of such an
event, a conclusion which holds even if the mass loss had been assumed to persist over
longer timescales (since the increased debris belt mass required would have been
recreated in collisions correspondingly less frequently).
Since $1M_\oplus$ is not an implausible remnant mass for an eccentric planetesimal belt 
(see \S \ref{ss:sd}), we consider secondary collisions to be a possible
explanation for HD69830, even if some process like sublimation must be invoked to
deplete small grains and render the collisional cascade of the parent planetesimal
belt non-detectable.

%%%%%%%%%%%%%%%%%%%%%%%%%%%%%%%%%%%%%%%%%%%%%%%%%%%%%%
%%%%%%%%%%%%%%%%%%%%%%%%%%%%%%%%%%%%%%%%%%%%%%%%%%%%%%
\section{Conclusion}
\label{s:conc}
This paper considers collisional processes in a population of
planetesimals with high eccentricities.
Collision rates are derived both analytically and using
Monte-Carlo simulations.
It was found that eccentricity has a significant effect on collision
rates, and that the amount of mass that can remain in a planetesimal belt
at late times can be significantly increased.

The emission properties of eccentric planetesimal belts were presented, and it
was found that radiation pressure causes eccentric rings to be deficient in small 
particles.
Thus, despite the increased mass of a high eccentricity planetesimal belt
at late times, extreme eccentricities of $>0.99$ are required to enhance the emission level
above that expected from a low eccentricity belt.
However, the high mass loss rate of extreme eccentricity planetesimal belts can
cause the wind of blow-out particles that extends outward from the pericentre to
be detectable.
The high frequency of massive collisions in such belts can also make it 
likely for us to be witnessing dust produced in such collisions.

Application of this model to the $\eta$ Corvi debris disk showed that all
available observations can be explained by an extreme eccentricity ($e=0.99$)
planetesimal belt of mass $5M_\oplus$, circumventing the conclusion that the hot
dust at 1.7AU must be transient.
Despite this success, the dynamical challenges of creating such a massive extreme
eccentricity population would need to be overcome before this model can be considered
viable.
Observational tests are suggested including the presence of resolvable emission 
(and absence of planets) in the 3-100AU region.
Application to HD69830 is complicated by the lack of far-infrared emission.
It may be possible to explain this non-detection by the removal of small
dust from the collisional cascade by sublimation, in which case the mid-infrared
emission may be plausibly explained by the ongoing destruction of a debris belt produced
in a recent collision in an eccentric planetesimal belt.

Although the majority of the discussion focuses on single eccentricity
populations, the results can also be applied to populations
with a range of semimajor axes and eccentricities.
This was demonstrated by application to scattered disk-like populations where it was 
found that, in the absence of other dynamical processes, rapid collisional erosion of 
the low eccentricity populations would be expected to result in a single high 
eccentricity population.
Since the known high eccentricity planetesimal populations are produced in
interactions with planets, and so may be continuing to undergo dynamical
evolution on timescales shorter than collisional timescales, it is noted 
that dynamical interactions may continue to play a defining role in the 
long term evolution of high eccentricity populations, and that the 
collisional evolution scheme presented here could be readily incorporated 
into N-body simulations of planet-planetesimal interactions to derive 
simultaneously the collisional and dynamical evolution of a scattered 
planetesimal population.
A further extension of the model would include a prescription for planetesimal
strength as a function of size which would lead to a departure from the single
phase size distribution assumed here (e.g., L\"{o}hne et al. 2008).

The results of this study would be applicable wherever non-negligible planetesimal
eccentricities are found.
Thus, other potential applications include the Solar System's comet and 
NEO populations, particularly in the early phases when these populations would have
been more massive and hence collisional processing more important (Booth et al. 2009),
the outcome of planet formation models (e.g., Payne et al. 2009), and
systems where eccentric planetesimals may be implicated such as the origin
of dust around White Dwarfs (Farihi, Jura \& Zuckerman 2009) and of the
hottest dust population of debris disks like Vega (Absil et al. 2006).

%%%%%%%%%%%%%%%%%%%%%%%%%%%%%%%%%%%%%%%%%%%%%%%%%%%%%%
%%%%%%%%%%%%%%%%%%%%%%%%%%%%%%%%%%%%%%%%%%%%%%%%%%%%%%
\section*{Acknowledgments}
The authors are grateful to the Isaac Newton Institute for Mathematical Sciences in Cambridge
where the final stages of this work were carried out during the Dynamics of Discs and Planets
research programme.

%%%%%%%%%%%%%%%%%%%%%%%%%%%%%%%%%%%%%%%%%%%%%%%%%%%%%%


\begin{thebibliography}{}

%% Reference example %%%%%%%%%%%%%%%%%

\bibitem[]{absil06} 
  Absil O., et al., 2006, A\&A, 452, 237
\bibitem[]{akeson09}
  Akeson R. L., et al., 2009, ApJ, 691, 1896 
\bibitem[]{alibert06} 
  Alibert Y., et al., 2006, A\&A, 455, L25
\bibitem[]{backman09}
  Backman D., et al., 2009, ApJ, 690, 1522
\bibitem[]{beichman05}
  Beichman C. A., Bryden G., Gautier T. N., Stapelfeldt K. R., Werner M. W., Misselt 
  K., Rieke G., Stansberry J., Trilling D., 2005, ApJ, 626, 1061
\bibitem[]{benz99}
  Benz W., Asphaug E., 1999, Icarus, 142, 5
\bibitem[]{booth09}
  Booth M., Wyatt M. C., Morbidelli A., Moro-Mart\'{i}n A., Levison H. F., 2009,
  MNRAS, in press
\bibitem[]{bottke93}
  Bottke W. F., Greenberg R., 1993, Geophys. Res. Lett., 20, 879
\bibitem[]{bottke94}
  Bottke W. F., Nolan M. C., Greenberg R., Kolvoord R. A., 1994, Icarus, 107, 255
\bibitem[]{bottke02}
  Bottke W. F., Morbidelli A., Jedicke R., Petit J.-M., Levison H. F.,
  Michel P., Metcalfe T. S., 2002, Icarus, 156, 399
\bibitem[]{bottke05}
  Bottke W. F., Durda D. D., Nesvorny D., Jedicke R., Morbidelli A.,
  Vokrouhlicky D., Levison H., 2005, Icarus, 175, 111
\bibitem[]{campo94}
  Campo-Bagatin A., Cellino A., Davis D. R., Farinella P., Paolicchi P., 1994,
  Planet. Space Sci., 42, 1079
\bibitem[]{chen06}
  Chen C. H., Sargent B. A., Bohac C., Kim K. H., Leibensperger E., Jura M.,
  Najita J., Forrest W. J., Watson D. M., Sloan G. C., Keller L. D., 2006, ApJSS, 
  166, 351
\bibitem[]{chen09}
  Chen C. H., Sheehan P., Watson D. M., Manoj P., Najita J. R.,
  2009, ApJ, 701, 1367
\bibitem[]{davis89}
  Davis D. R., Weidenschilling S. J., Farinella P., Paolicchi P., Binzel R. P., 1989,
  in Binzel R. P., Gehrels T., Matthews M. S., eds, Asteroids II.
  Univ. of Arizona Press, Tucson, p. 805
\bibitem[]{davis97}
  Davis D. R., Farinella P., 1997, Icarus, 125, 50
\bibitem[]{delloro98}
  Dell'Oro A., Paolicchi P., 1998, Icarus, 136, 328
\bibitem[]{difolco07}
  di Folco E., et al., 2007, A\&A, 475, 243
\bibitem[]{Dohnanyi69}
  Dohnanyi J. S., 1969, J. Geophys. Res., 74, 2531
\bibitem[]{dominik03}
  Dominik C., Decin G., 2003, ApJ, 598, 626
\bibitem[]{durda98}
  Durda D. D., Greenberg R. R., Jedicke R., 1998, Icarus, 135, 431
\bibitem[]{duncan08}
  Duncan, M. J., 2008, Sp. Sci. Rev., 138, 109
\bibitem[]{edgar04}
  Edgar R., Artymowicz P., 2004, MNRAS, 354, 769
\bibitem[]{farihi09}
  Farihi J., Jura M., Zuckerman B., 2009, ApJ, 694, 805
\bibitem[]{fujiwara89}
  Fujiwara A., Cerroni P., Davis D. R., Ryan E., di Martino M., Holsapple K.,
  Housen K., 1989, in Binzel R. P., Gehrels T., Matthews M. S., eds, Asteroids II.
  Univ. of Arizona Press, Tucson, p. 240
\bibitem[]{gomes05}
  Gomes R., Levison H. F., Tsiganis K., Morbidelli A., 2005, Nature, 435, 466
\bibitem[]{gomes08}
  Gomes R. S., Fern\'{a}ndez J. A., Gallardo T., Brunini A., 2008. In The Solar System
  Beyond Neptune, ed. A. Barucci et al. (Tucson: Univ. Arizona Press), 259
\bibitem[]{greaves05}
  Greaves J. S., et al., 2005, ApJ, 619, L187
\bibitem[]{greenberg78}
  Greenberg R., Hartmann W. K., Chapman C. R., Wacker J. F., 1978, Icarus, 35, 1
\bibitem[]{greenberg82}
  Greenberg R., 1982, AJ, 87, 184
\bibitem[]{greenzweig90}
  Greenzweig Y., Lissauer J. J., 1990, Icarus, 87, 40
\bibitem[]{hartmann00}
  Hartmann W. K., Ryder G., Dones L., Grinspoon D., 2000, in
  Canup R. M., Righter K., eds., Origin of the Earth and Moon.
  Univ. of Ariz. Press, Tucson, p. 493
\bibitem[]{heng09}
  Heng K., Tremaine S. 2009, MNRAS, in press, arXiv:0909.3850
\bibitem[]{ipatov08}
  Ipatov S. I., Kutyrev A. S., Madsen G. J., Mather J. C. Moseley S. H.,
  Reynolds R. J., 2008, Icarus, 194, 769
\bibitem[]{kalas05}
  Kalas P., Graham J. R., Clampin M., 2005, Nature, 435, 1067
\bibitem[]{kenyon04}
  Kenyon S. J., Bromley B. C., 2004, AJ, 128, 1916
\bibitem[]{kobayashi09}
  Kobayashi H., Tanaka H., 2009, in press
\bibitem[]{krivov05}
  Krivov A. V., Sremcevic M., Spahn F., 2005, Icarus, 174, 105
\bibitem[]{krivov06}
  Krivov A. V., L\"{o}hne T., Sremcevic M., 2006, A\&A, 455, 509
\bibitem[]{lagrange09}   
  Lagrange A. M., Desort M., Galland F., Udry S., Mayor M., 2009, A\&A, 495, 335 
\bibitem[]{leinhardt09}   
  Leinhardt Z. M., Stewart S. T., 2009, Icarus, 199, 542
\bibitem[]{levison08}   
  Levison H. F., Morbidelli A., Vokrouhlicky D., Bottke W. F., 2008, AJ, 136, 1079
\bibitem[]{li98}
  Li A., Greenberg J. M., 1998, A\&A, 331, 291
\bibitem[]{lissauer93}   
  Lissauer J. J., 1993, ARA\&A, 31, 129
\bibitem[]{lissauer93b}   
  Lissauer J. J., Stewart G. R., 1993, in Levy E. H. \& Lunine J. I., eds,
  Protostars \& Planets III. Univ. of Arizona Press, Tucson, p. 1061
\bibitem[]{lisse07}
  Lisse C. M., Beichman C. A., Bryden G., Wyatt M. C., 2007, ApJ, 658, 584
\bibitem[]{lohne07}
  L\"{o}hne T., Krivov A. V., Rodmann J., 2008, ApJ, 673, 1123
\bibitem[]{lovis06}
  Lovis C., et al., 2006, Nature, 441, 305
\bibitem[]{mandell07}
  Mandell A. M., Raymond S. N., Sigurdsson S., 2007, ApJ, 660, 823
\bibitem[]{mathhews07a}
  Matthews B. C., Kalas P. G., Wyatt M. C., 2007, ApJ, 663, 1103
\bibitem[]{moor09}
  Mo\'{o}r A., Apai D., Pascucci I., \'{A}brah\'{a}m P, Grady C., Henning Th., Juh\'{a}sz A.,
  Kiss Cs., K\'{o}sp\'{a}l \'{A}, 2009, ApJL, 700, L25
\bibitem[]{moromartin08}
  Moro-Mart\'{i}n A., Wyatt M. C., Malhotra R., Trilling D. E., 2008.
  In Barucci A. et al., eds, The Solar System Beyond Neptune.
  Univ. of Arizona Press, Tucson, p. 465
\bibitem[]{nesvorny09}
  Nesvorn\'{y} D., Jenniskens P., Levison H. F., Bottke W. F., Vokrouhlick\'{y} D.,
  2009, in press, arXiv:0909.4322
\bibitem[]{obrien03}
  O'Brien D. P., Greenberg R., 2003, Icarus, 164, 334
\bibitem[]{obrien05}
  O'Brien D. P., Greenberg R., 2005, Icarus, 178, 179
\bibitem[]{opik51}
  \"{O}pik E. J., 1951, Proc. R. Irish Acad., 54, 165
\bibitem[]{payne09}
  Payne M. J., Ford E. B., Wyatt M. C., Booth M., 2009, MNRAS, 393, 1219
\bibitem[]{reach09}
  Reach W. T., Vaubaillon J., Kelley M. S., Lisse C. M., Sykes M. V., 2009,
  Icarus, 203, 571
\bibitem[]{schneider09}
  Schneider G., Weinberger A. J., Becklin E. E., Debes J. H., Smith B. A.,
  2009, AJ, 137, 53
\bibitem[]{sheret04}
  Sheret I., Dent W. R. F., Wyatt M. C., 2004, MNRAS, 348, 1282
\bibitem[]{smith08}
  Smith R., Wyatt M. C., Dent W. R. F., 2008, A\&A, 485, 897
\bibitem[]{smith09a}
  Smith R., Churcher L. J., Wyatt M. C., Moerchen M. M., Telesco C. M.,
  2009a, A\&A, 493, 299
\bibitem[]{smith09b}
  Smith R., Wyatt M. C., Haniff C. A., 2009b, A\&A, 503, 265
\bibitem[]{spaute91}
  Spaute D., Weidenschilling S. J., Davis D. R., Marzari F., 1991,
  Icarus, 92, 147
\bibitem[]{stern97}
  Stern S. A., Colwell J. E., 1997, ApJ, 490, 879
\bibitem[]{sykes90}
  Sykes M. V., 1990, Icarus, 85, 267
\bibitem[]{tanaka96}
  Tanaka H., Inaba S., Nakazawa K., 1996, Icarus, 123, 450
\bibitem[]{trilling07b}
  Trilling D. E., Bryden G., Beichman C. A., Rieke G. H., Su K. Y. L., Stansberry 
  J. A., Blaylock M., Stapelfeldt K. R., Beeman J. W., Haller E. E., 2008, ApJ,
  674, 1086
\bibitem[]{vedder96}
  Vedder J. D., 1996, Icarus, 123, 436
\bibitem[]{volk08}
  Volk K., Malhotra R., 2008, ApJ, 687, 714
\bibitem[]{wetherill67}
  Wetherill G. W., 1967, J. Geophys. Res., 72, 2429
\bibitem[]{wetherill89}
  Wetherill G. W., Stewart G. R., 1989, Icarus, 77, 330
\bibitem[]{wyatt02}
  Wyatt M. C., Dent W. R. F., 2002, MNRAS, 334, 589
\bibitem[]{wyatt99}
  Wyatt M. C., Dermott S. F., Telesco C. M., Fisher R. S., Grogan K., Holmes E. K., 
  Pi\~{n}a R. K., 1999, ApJ, 527, 918
\bibitem[]{wyatt05c}
  Wyatt M. C., Greaves J. S., Dent W. R. F., Coulson I. M., 2005,
  ApJ, 620, 492  
\bibitem[]{wyatt07a}
  Wyatt M. C., Smith R., Beichman C. A., Bryden G., Greaves J. S., Lisse C. M., 
  2007a, ApJ, 658, 569
\bibitem[]{wyatt07b}
  Wyatt M. C., Smith R., Su K. Y. L., Rieke G. H., Greaves J. S., Beichman C. A., 
  Bryden G., 2007b, ApJ, 663, 365
\bibitem[]{wyatt08} Wyatt M. C., 2008, ARA\&A, 46, 339
\bibitem[]{wyatt09} Wyatt M. C., 2009, Lect. Notes Phys., 758, 37

\end{thebibliography}
\end{document}